# VERIFICATION OF MODELS FOR CALCULATION OF E1 RADIATIVE STRENGTH

**Vladimir A. Plujko**[1]

*Nuclear Physics Department, Taras Shevchenko National University*
*Prospekt Akad. Glushkova, 2, k.11, 03022 Kyiv, Ukraine*
*E-mail:*`plujko@univ.kiev.ua`

**Igor M. Kadenko**

*Nuclear Physics Department, Taras Shevchenko National University*
*Prospekt Akad. Glushkova, 2, k.11, 03022 Kyiv, Ukraine*
*E-mail:* `ndef@uninet.kiev.ua`

**Elizaveta V. Kulich**

*Nuclear Physics Department, Taras Shevchenko National University*
*Prospekt Akad. Glushkova, 2, k.11, 03022 Kyiv, Ukraine*
*E-mail:* `lizak@univ.kiev.ua`

**Stephane Goriely**

*Institut d'Astronomie et Astrophysique Université Libre de Bruxelles*
*Brussels, Belgium*
*E-mail:* `sgoriely@astro.ulb.ac.be`

**Oksana I. Davidovskaya**

*Kyiv Institute for Nuclear Research*
*Prospekt Nauki, 47, 03680 Kyiv, Ukraine*
*E-mail:* `oid@kinr.kiev.ua`

**Oleksandr M. Gorbachenko**

*Nuclear Physics Department, Taras Shevchenko National University*
*Prospekt Akad. Glushkova, 2, k.11, 03022 Kyiv, Ukraine*
*E-mail:* `gorbachenko@univ.kiev.ua`

Photoabsorption cross sections and γ-decay strength function are calculated and compared with experimental data to test the existing models of dipole radiative strength functions (RSF) for the middle-weight and heavy atomic nuclei. Simplified version of the modified Lorentzian model are proposed. New tables of giant dipole resonance (GDR) parameters are given. It is shown that the phenomenological closed-form models with asymmetric shape can be used for overall estimates of the dipole RSF in the γ-ray energy region up to about 20 MeV when GDR parameters are known or the GDR systematics can be adopted. Otherwise, the HFB-QRPA microscopic model and the semi-classical approach with moving surface appear to be more adequate methods to estimate the dipole photoabsorption RSF.

---

[1] Speaker

## 1. Introduction

Gamma-emission is one of the most universal channels of the nuclear de-excitation processes which accompanies any nuclear reaction. The average probability for a γ-transition can be described through the use of the radiative strength functions (RSF)[1,2]. Both the $\gamma$ -ray emission and absorption processes are connected. Their respective radiative strengths are usually denoted as $\overleftarrow{f}_{XL}$ and $\overrightarrow{f}_{XL}$ for γ-transitions of electric (*X*=*E*) or magnetic (*X*=*M*) type with multipolarity $L$ .

Dipole electric γ-transitions ($E1$) are dominant, when they occur simultaneously with transitions of other multipolarities and types. Therefore we focus here on the dipole RSF. The average dipole radiative width $\overline{\Gamma}_{E1}$ per unit of the $\gamma$ - ray energy interval is determined by a γ-decay (downward) strength function $\overleftarrow{f}_{E1}$ in the following way

$$\overleftarrow{f}_{E1}\left(E_\gamma\right)=\frac{\overline{\Gamma}_{E1}\left(E_\gamma\right)}{3E_\gamma^3}\frac{\rho\left(U\right)}{\rho\left(U-E_\gamma\right)}=F_{E1}(E_\gamma,T_f)\,, \tag{1}$$

where $\rho\left(U\right)$ is the total density of the excited states in heated nuclei at initial excitation energy $U$ (initial temperature $T$ ); $E_\gamma$ is the γ-ray energy and $T_f$ the final state temperature.

The photoexcitation (upward) strength function $\overrightarrow{f}_{E1}$ is connected to the total photoabsorption cross-section $\sigma_{E1}$ by

$$\overrightarrow{f}_{E1}\left(E_\gamma\right)=\frac{\sigma_{E1}\left(E_\gamma\right)}{3E_\gamma\left(\pi\hbar c\right)^2}=F_{E1}(E_\gamma,T)\,. \tag{2}$$

The radiative strengths of the γ-decay and the photoabsorption for heated nuclei are described in term of a spectral function $F_{E1}$ which is defined by the similar expression as for the de-excitation channel but with different temperature. More specifically, the $\gamma$ -decay strength function depends on the temperature $T_f$ of the final state. This temperature is a function of the $\gamma$ - ray energy in contrast to the initial state temperature $T$ .

According to general relation between the γ-decay RSF of heated nuclei and the imaginary part of the response function to an electromagnetic field ([4-8]), the spectral function $F_{XL}(E_\gamma,\tau)$ (with $\tau$ standing for $T$ or $T_f$ ) is proportional to the strength function $S_{XL}$ of the nuclear response to the electromagnetic field of type $XL$ with frequency $\omega=E_\gamma/\hbar$ :

$$F_{XL}(E_\gamma,\tau)=\frac{4\pi}{9}\frac{e^2}{(\hbar c)^3}\cdot\Lambda(E_\gamma,\tau)\cdot S_{XL}(\omega),\qquad \Lambda(E_\gamma,\tau)=\frac{1}{1-\exp(-E_\gamma/\tau)}\,. \tag{3}$$

The scaling factor $\Lambda(E_\gamma,\tau)$ determines the enhancement magnitude of the radiative strength functions in heated nuclei at the temperature $\tau$ as compared to the zero temperature case. It can be considered as the average number of 1p-1h states, $N_{1p-1h}$, excited by the electromagnetic field:

$$\Lambda(E_\gamma,\tau) = N_{1p-1h} \equiv \int_0^{+\infty} \frac{d\varepsilon}{E_\gamma} \overline{f}(\varepsilon)(1-\overline{f}(\varepsilon+E_\gamma)) . \qquad (4)$$

Here, $\overline{f}(\varepsilon) = 1/[1+\exp((\varepsilon-\mu)/\tau)]$ is the equilibrium distribution of the single-particle states at the temperature $\tau$ and $\mu$ the corresponding chemical potential (which equals the Fermi energy $\varepsilon_F$ for $\tau << \varepsilon_F$ ).

The strength function $S_{XL}$ of the nuclear response to the electric field with multipolarity $L$ is proportional to the imaginary (dissipative) part $\chi''_{XL}$ of the nuclear response function $\chi_{XL}$ to electromagnetic field with multipolarity $L$:

$$S_{XL}(\omega) \equiv -\frac{1}{\pi}\chi''_{XL}(\omega) , \qquad (5)$$

where in the case of electric transitions in spherical nuclei, the external potential has the form

$$V_{ext}^L = q_\omega(t) Q_L(r) Y_{L0}(\hat{r}),\ \ q_\omega(t) = q_0 \exp[-i(\omega+i\delta)t],\ \ \delta \to +0,\ \ q_0 << 1 , \qquad (6)$$

with radial form factor $Q_L(r) = r^L$ .

It should be noted that the photoexcitation strength function for cold nuclei is determined by the spectral function (3) at $\tau = 0$, i.e. with $\Lambda \equiv 1$.

The RSF contains information on nuclear structure and they are auxiliary quantities involved in calculations of the observed characteristics of most nuclear reactions. Their microscopic determinations, as a rule, are time-consuming and for this reason, simple closed-form expressions are often preferable for their evaluation.

In this contribution, different models of the RSF are tested. More specifically, experimental photoabsorption and γ-decay data are compared with theoretical calculations performed within the framework of the microscopic Hartree-Fock-Bogoliubov plus quasi-particle random phase approximation model (HFB-QRPA) [9,10], the semi-classical approach with moving surface (MSA) [11-13] and the more traditional Lorentzian-type models using closed-form expressions [1-3]. Here, we also consider the description of the dipole RSF within the simplified version of the generalized Lorentzian model in which the width is linearly dependent on energy.

## 2. Main features of the tested RSF models

Different models have been developed to describe the dipole RSF microscopically. We consider here the expressions for the linear response function of atomic nuclei to the electric dipole field within the framework of the HFB-QRPA approach from Refs. [9,10]. The semi-classical MSA method [11,12] is based on solving the kinetic Landau-Vlasov equation for finite systems with a moving surface [13].

Phenomenological models assume the dipole RSF to have a Lorentzian-like shape. Different expressions for the scaling parameter (or "width" $\Gamma_\gamma(E_\gamma)$) of the curve shape exist. The quantity $\Gamma_\gamma(E_\gamma)$ is governed by the damping of the collective states.

In the Standard Lorentzian model (SLO [14,15]), the Brink hypothesis is used to calculate the dipole γ-ray strength. The dipole RSF in the SLO model is a single Lorentzian (for spherical nuclei) with an energy-independent width $\Gamma_\gamma(E_\gamma)$ taken to be equal to the GDR width $\Gamma_r$:

$$\overleftarrow{f}\left(E_\gamma\right) = \vec{f}\left(E_\gamma\right) \equiv f_{SLO}\left(E_\gamma\right) = 8.674\cdot 10^{-8}\,\sigma_r\Gamma_r \frac{E_\gamma\Gamma_r}{\left(E_\gamma^2 - E_r^2\right)^2 + \left[\Gamma_r\cdot E_\gamma\right]^2}, \quad MeV^{-3}, \quad (7)$$

where the Lorentzian parameters $\sigma_r$, $E_r$ are the peak cross section and the GDR energy, respectively; the energies and width are expressed in units of *MeV*, and $\sigma_r$ in *mb*. A behavior of the width $\Gamma_\gamma(E_\gamma) = \Gamma_r$=*const* is similar to the fragmentation component of the collective state damping width that corresponds to redistribution of the $\gamma$-strength in a self-consistent mean-field when nucleon collisions in the nuclear interior are not taken into account. In the semiclassical approach, the fragmentation component of the damping width is governed by the nucleon collisions with a moving surface of the nucleus (one-body dissipation) and it is practically independent of the energy (see [16] and Refs. therein).

The SLO approach is probably the most appropriate method for describing photo-absorption data for medium-weight and heavy nuclei [1, 17, 18]. However, the SLO model for γ emission significantly underestimates the γ-decay spectra at low energies $E_\gamma \le 1\div 2$ *MeV* [19]. A global description of the γ-decay spectra by the Lorentzian approach can be obtained over the energy range $1\div 2 < E_\gamma < 8$ *MeV* but in that case the parameters become inconsistent with those derived from the photo-absorption data near the GDR peak. Generally, the SLO with the GDR parameters overestimates experimental data around the neutron separation energy, such as capture cross sections and the average radiative widths in heavy nuclei [1,20-23].

The first model with a correct description of the $E1$ strengths at energies $E_\gamma$ close to zero was developed by Kadmenskij, Markushev and Furman in Ref.[24] (KMF model). The KFM expression of the RSF in the limiting case of $E_\gamma$=0 is

$$\overleftarrow{f}(E_\gamma = 0) \equiv \overleftarrow{f}^{(0)}_{KMF} = 8.674\cdot 10^{-8}\cdot\sigma_r\cdot\Gamma_r\cdot K\cdot\frac{\Gamma_c(E_\gamma = 0)}{E_r^3}. \quad (8)$$

where the quantity $K$ is determined by the Landau parameters $F_0^1, F_1^1$ of the quasi-particle interaction in the isovector channel of the Fermi system:

$$K = \left(1 + F_1^1/3\right)^{1/2} \Big/ \left(1 + F_0^1/3\right)^{1/2} = \sqrt{E_r/E_0} \quad (9)$$

where $E_0$ is an average energy for the one-particle one-hole states forming the GDR. The value $K = K_{KMF} = 0.7$ is adopted in the KMF-model.

The damping parameter $\Gamma_c(E_\gamma)$ takes the form of the collisional component of the damping width of the zero sound in the infinite Fermi liquid.in which the GDR energy is replaced by γ–ray energy $E_\gamma$, i.e.

$$\Gamma_c(E_\gamma) = C_{coll}\left(E_\gamma^2 + 4\pi T_f^2\right). \qquad (10)$$

It should be mentioned that according to a semi-classical approach based on the Landau-Vlasov equation, the energy dependence of the collisional width $\Gamma_c(E_\gamma)$ results from the non-Markovian form of the collision integral by allowing for retardation effects (Appendix A).

The constant $C_{coll}$ in Eq.(10) is derived from the normalization condition of $\Gamma_c(E_\gamma)$, with respect to the GDR width in cold nuclei:

$$C_{coll} = \frac{\Gamma_r}{E_r^2} \equiv C_{KFM}\,. \qquad (11)$$

This descrpition is one of the foundation of the Enhanced Generalized Lorentzian (EGLO) model [2,21,25] as well as the Generalized Fermi-Liquid (GFL) model[26] (with the extension [27] that are used to describe the E1 γ-decay RSF in the range of γ-ray energies up to the GDR energy).

For spherical nuclei, the EGLO RSF consists of two components: (i) a Lorentzian with an energy- and temperature-dependent empirical width, and (ii) a term of the shape corresponding to the KMF limit (8) of the RSF at zero value of the γ-ray energy:

$$\overleftarrow{f}(E_\gamma) \equiv \overleftarrow{f}_{EGLO}(E_\gamma) = 8.674\cdot 10^{-8}\cdot\sigma_r\Gamma_r\left[\frac{E_\gamma\Gamma_K(E_\gamma)}{(E_\gamma^2-E_r^2)^2+E_\gamma^2\Gamma_K^2(E_\gamma)} + \frac{0.7\Gamma_K(E_\gamma=0)}{E_r^3}\right]. \qquad (12)$$

The EGLO-width has form of the expression (10) but with the energy-dependent parameter $C_{coll}$:

$$\Gamma_K(E_\gamma) = C_{coll}(E_\gamma)\cdot\left(E_\gamma^2 + 4\pi T_f^2\right),$$

$$C_{coll}(E_\gamma) = \frac{\Gamma_r}{E_r^2}\cdot\chi(E_\gamma) = C_{KMF}\,\chi(E_\gamma) \equiv C_K(E_\gamma)\,, \qquad (13)$$

where $\chi(E_\gamma)$ is a function obtained from a fit to experimental data,

$$\chi(E_\gamma) = k + (1-k)(E_\gamma - \varepsilon_0)/(E_r - \varepsilon_0)\ , \qquad (14)$$

where the parameter $k$ reproduces the experimental E1 strength around a reference energy $\varepsilon_0$; $\chi(E_\gamma = E_r) = 1$ and $C_K(E_r) = C_{KMF}$. The value of the factor $k$ depends on the model adopted to describe the nuclear state density, and was obtained from the average resonance capture data, while $\varepsilon_0 = 4.5\ MeV$. If the Fermi-gas model is used, $k$ is given by [2]

$$k = \begin{cases} 1, & A < 148, \\ 1 + 0.09\ (\text{A-148})^2 \exp\ \{\text{-0.18 (A-148)}\}\ , & A \geq 148. \end{cases} \qquad (15)$$

The value $\Gamma_K(E_\gamma = 0)$ in Eq.(13) is equal to $\chi(0)\cdot\Gamma_r\cdot 4\pi T^2/E_r^2$.

The expression for dipole γ-decay RSF within the GFL model has the following form[3,26,27]

$$\overleftarrow{f}(E_\gamma) \equiv \overleftarrow{f}_{GFL}(E_\gamma) = 8.674 \cdot 10^{-8} \cdot \sigma_r \Gamma_r \frac{K \cdot E_r \cdot \Gamma_m(E_\gamma)}{\left(E_\gamma^2 - E_r^2\right)^2 + K\left[\Gamma_m(E_\gamma) E_\gamma\right]^2}, \qquad (16)$$

This equation is an extension [27] of the original expression [26], in which the term $K$ has been added to the denominator to avoid a singularity in the GFL approach near the GDR energy. The quantity $K$ is determined by Eq.(9) but the value $K = K_{GFL} = 0.63$ is adopted. The energy-dependent width $\Gamma_m(E_\gamma)$ is taken to be a sum of a collisional damping width $\Gamma_{coll}$ and the additional term $\Gamma_{dq}$:

$$\Gamma_m(E_\gamma) = \Gamma_{coll}\left(E_\gamma\right) + \Gamma_{dq}\left(E_\gamma\right) . \qquad (17)$$

The collisional component corresponds to an extension (10) of the zero sound damping width in the infinite Fermi-liquid model

$$\Gamma_{coll}(E_\gamma) \equiv C_{coll}\left(E_\gamma^2 + 4\pi^2 T_f^2\right) , \qquad (18)$$

but with the constant parameter $C_{coll}$ which is determined by normalizing the total width (17) at $E_\gamma = E_r$ and $T_f = 0$ to the GDR width of a cold nucleus, i.e. $\Gamma_m(E_\gamma = E_r) = \Gamma_r$. The component $\Gamma_{dq}$ results from the damping of the nuclear response due to quadrupole vibrations of the nuclear surface. It is taken in the form of the first term of Eq.(16) from Ref.[28] for the spreading width of the GDR over the surface quadrupole vibrations by replacing the GDR energy by the γ–ray energy $E_\gamma$:

$$\Gamma_{dq}\left(E_\gamma\right) = C_{dq}\sqrt{E_\gamma^2 \overline{\beta}_2^2 + E_\gamma s_2} , \qquad (19)$$

where $C_{dq} = \sqrt{5 \ln 2/\pi} = 1.05$; $s_2 = E_{2^+}\overline{\beta}_2^2 \approx 217.16/A^2$ with $E_{2^+}$ being the energy of the first vibrational quadrupole state, and $\overline{\beta}_2$ is the effective deformation parameter characterizing the nuclear stiffness with respect to surface vibrations.

The photoexcitation strength function $\overrightarrow{f}_{E1}$ can also be calculated within the EGLO and GFL models. The corresponding expressions $\overrightarrow{f}_{E1}$ in cold nuclei are determined by Eqs.(12), (16) and (17), (18) with $T_f = 0$.

It should be mentioned that the low-energy behavior of the γ -decay RSF within the KMF, EGLO and GFL models are similar $\overleftarrow{f}_{E1}(E_\gamma \to 0) \to const$ but the values of $\overleftarrow{f}_{E1}(0)$ are different due to differences of both the contributions of the temperature-dependent component in the widths $\Gamma_c(E_\gamma)$, $\Gamma_K(E_\gamma)$, $\Gamma_m(E_\gamma)$ at $E_\gamma = 0$ and the adopted values for the Landau parameters of the quasi-particle interaction:

$$\overleftarrow{f}_{\alpha}(E_{\gamma}=0)=8.674\cdot 10^{-8}\cdot\sigma_r\Gamma_r\frac{K_{\alpha}\cdot\Gamma_{\alpha}(E_{\gamma}=0)}{E_r^3}\quad(MeV^{-3})\,, \qquad (20)$$

where the index $\alpha$ denotes the corresponding model.

It can be noted that the SLO, EGLO and GFL expressions for the γ-decay strength function of heated nuclei are in fact usual parameterizations of experimental data. They are in contradiction with some aspects of the microscopic theoretical studies. Specifically, the shapes of the radiative strengths within these approaches are not consistent with the general relations (1), (3)-(6) between a RSF and the imaginary part of the response function. To avoid this shortcoming, at least approximately, an approach was proposed in Ref.[29] which was named later in [3,30] as the Modified Lorentzian approach (MLO). It is based on general equations between the RSF and the imaginary part of the nuclear response function with the Lorentzian-like shape and an energy-dependent broadening parameter (width) $\Gamma_{\gamma}$

$$\operatorname{Im}\chi\left(\omega=E_{\gamma}/\hbar\right)\propto\frac{E_{\gamma}\Gamma_{\gamma}\left(E_{\gamma}\right)}{\left(E_{\gamma}^2-E_r^2\right)^2+\left[\Gamma_{\gamma}\left(E_{\gamma}\right)E_{\gamma}\right]^2}\,. \qquad (21)$$

This shape results from a semi-classical approach based on the Landau-Vlasov equation with a non-Markovian collision term if the γ-transition strength is concentrated near the giant resonance (see [30] and Appendix A). The non-Markovian form of the collision integral allowing for retardation effects of the nucleon-nucleon collision just leads to the energy dependence of the width $\Gamma_{\gamma}$.

The Lorentzian shape also stems from the random-phase approximation in cold nuclei [31] as well as from the extended hydrodynamic model of Steinwedel-Jensen [8] for heated nuclei with friction between the proton and neutron fluids.

As a consequence, the γ-decay RSF within the framework of the MLO model has the following form,

$$\overleftarrow{f}\left(E_{\gamma}\right)=\overleftarrow{f}_{MLO}\left(E_{\gamma}\right)=8.674\cdot 10^{-8}\,\sigma_r\Gamma_r\frac{E_{\gamma}}{1-\exp\left(-E_{\gamma}/T_f\right)}\frac{\Gamma_{\gamma}(E_{\gamma})}{\left(E_{\gamma}^2-E_r^2\right)^2+\left[\Gamma_{\gamma}(E_{\gamma})\cdot E_{\gamma}\right]^2}\,, \qquad (22)$$

where $\Gamma_r=\Gamma_{\gamma}(E_{\gamma}=E_r)$ at zero excitation energy; the width $\Gamma_{\gamma}(E_{\gamma})$ depends on the assumptions on the damping mechanism for the collective states (see [3,27,30] for details and references).

Different semi-empirical expressions for the width were previously used in the MLO approach (MLO1, MLO2, MLO3) but, as a rule, the resulting RSF are in rather close agreement. The MLO1 model is based on the semi-classical approach with the use of the Landau-Vlasov equation with a collision term [32] (see Appendix A). As a result, the width can be expressed as:

$$\Gamma_{\gamma}\left(E_{\gamma}\right)=\hbar\,\frac{\beta(E_{\gamma};\tau_c(E_{\gamma},U_f))}{\tau_c(E_{\gamma},U_f)}\,, \qquad (23)$$

where the function $\beta(E_\gamma;\tau_c(E_\gamma,U_f))$, Eq.(A27), is almost constant in the $\gamma$–ray energy range that sligthly exceeds the GDR energy; $U_f$ is the excitation energy of the final state; $\tau_c(E_\gamma=\hbar\omega;V=0)$ the collisional relaxation time of the collective motion in cold Fermi system under an external field with frequency $\omega$. At $E_\gamma=E_r$, $\tau_c(E_\gamma=E_r;V=0)$ is the collisional relaxation time of the GDR.

The dependence of $\tau_c(E_\gamma=\hbar\omega;V=0)$ on the γ-ray energy results from retardation (memory) effects in the collision integral which is used to describre the scattering particles in the nuclear interior [33,34]. In the MLO1 approach, the relaxation time is found by analogy with the relaxation times of states in the exciton nuclear reaction model and it is taken as ([3,33])

$$\frac{\hbar}{\tau_c\left(E_\gamma;V\right)}=b\left(E_\gamma+V\right), \tag{24}$$

where $b$ is a constant determined by in-medium cross section of neutron-proton scattering. The magnitude of $b$ is determined from the condition $\Gamma_r=\Gamma_\gamma(E_\gamma=E_r)$ in cold nuclei, i.e. from the relationship $\beta(E_r;\tau_c(E_r,V=0))/\tau_c(E_r,V=0)=\Gamma_r/\hbar$. A linear energy dependence of the collisional relaxation time in Eq.(24) results from the inverse proportionality of the effective mean square matrix element to the excitation energy for transitions between the exciton states [35]. The linear energy dependence of the collisional width (which is proportional to $\hbar/\tau_c$) was also obtained within the test particle approach, when nucleon collisions were considered as *s*-wave scattering between pseudo-particles [36,37].

The MLO strength function $\vec{f}_{E1}$ describing the photoexcitation of cold nuclei is given by the expression

$$\vec{f}\left(E_\gamma\right)=\vec{f}_{MLO}\left(E_\gamma\right)=8.674\cdot10^{-8}\,\sigma_r\Gamma_r\frac{E_\gamma\Gamma_\gamma(E_\gamma)}{\left(E_\gamma^2-E_r^2\right)^2+\left[\Gamma_\gamma(E_\gamma)\cdot E_\gamma\right]^2} \tag{25}$$

with $\Gamma_\gamma(E_\gamma)$ of MLO1 model is determined by Eq.(23) with the following relaxation time:

$$\Gamma_\gamma\left(E_\gamma\right)=\beta(E_\gamma;\tau_c(E_\gamma,V=0))\cdot\frac{\hbar}{\tau_c(E_\gamma,V=0)},\quad \frac{\hbar}{\tau_c\left(E_\gamma;V=0\right)}=b\,E_\gamma\,. \tag{26}$$

It can be noted that, in contrast to Eq. (26), the relaxation time (24) (used for calculation of the the γ–decay strengh function within MLO1) is in fact independent of the γ-ray energy and depends only on the initial excitation energy $U=E_\gamma+U_f$.

In the case the excitation energy is not too high and $E_\gamma$ ranges from zero up to the GDR energy, the function $\beta$ in Eq.(23) depends only weakly on the energy and the MLO1 width $\Gamma_\gamma\left(E_\gamma\right)$ can be expressed as

$$\Gamma_\gamma\left(E_\gamma\right)=\begin{cases} a(E_\gamma+U_f)=aU, & \text{for } \gamma-\text{decay}, \\ a\,E_\gamma\,, & \text{for photoabsorption}, \end{cases} \tag{27}$$

where $a=\Gamma_r/E_r=C_{KMF}\cdot E_r$ if the normalization condition $\Gamma_r=\Gamma_\gamma(E_\gamma=E_r)$ is adopted for cold nuclei.

Below we also test the RSF description within the modified Lorentzian model given by Eqs.(22), (25) with the simplified expression (27) for $\Gamma_\gamma\left(E_\gamma\right)$. This model is denoted as the Simplified Modified Lorentzian (SMLO) model. The valuesof the parameter $a$ is obtained by fitting the SMLO shape to the experimental photoabsorption cross sections (Eqs.(2),(25),(27)) in spherical nuclei .Results are presented in Appendix B.

The RSF of the MLO2 and MLO3 models also takes the form of Eq.(25) but an approximation of independent dissipation sources for the widths[ 16] is considered. The widths are taken as a sum of the collisional component $\hbar/\tau_c$ and a term $k_s\,\hbar/\tau_w$ which simulate the fragmentation contribution to the width:

$$\Gamma_\gamma\left(E_\gamma\right)=\frac{\hbar}{\tau_c\left(E_\gamma\right)}+k_s(E_\gamma)\frac{\hbar}{\tau_w}. \tag{28}$$

The collisional relaxation time is determined by Eq.(24) as in the case of the MLO2 model. In the MLO3 model, the collective relaxation time reads

$$\frac{\hbar}{\tau_c\left(E_\gamma\right)}=\Gamma_{coll}(E_\gamma)=C_{coll}\left(E_\gamma^2+4\pi^2T_f^2\right). \tag{29}$$

Here, a magnitude of $C_{coll}$ is detemined by the neutron-proton cross section $\sigma_{in}(n,p)$ in the nuclear medium near the Fermi surface

$$C_{coll}=F\cdot c,\quad F=\frac{\sigma_{in}(n,p)}{\sigma_{free}(n,p)},\quad c=\frac{4}{9\pi^2}\frac{m}{\hbar^2}\sigma_{free}(n,p)=5.386\cdot10^{-3}\ (MeV^{-1}), \tag{30}$$

where the value $\sigma_{free}(n,p)=5\ fm^2$ is adopted for the free space cross section near the Fermi surface and $\hbar^2/m=\ 41.80349\ (MeV\ fm^2)$ .

The expression (29) is obtained within framework of the kinetic theory description of the nuclear excitations by including retardation effects in the collision integral (see Appendix A for references).

The magnitude of the second (fragmentation) component in Eq.(28) is taken proportional to the wall formula value [38]

$$\frac{\hbar}{\tau_w}=\Gamma_w=\frac{3}{4}\frac{\hbar v_F}{R_0}=\frac{32.846}{A^{1/3}}\ (MeV) \tag{31}$$

with the magnitude $\varepsilon_F=mv_F^2/2=37\ MeV$ for the Fermi energy and $R_0=r_0A^{1/3}$ with $r_0=1.27\ fm$. A scaling factor $k_s$ in Eq.(28) is included in the energy-dependent power approximation:

$$k_s\left(E_\gamma\right)=\begin{cases}k_r+\left(k_0-k_r\right)\left|\left(E_\gamma-E_r\right)/E_r\right|^{n_s}, & E_\gamma<2E_r,\\ k_0, & E_\gamma\geq 2E_r,\end{cases} \quad (32)$$

where the constant $k_r$ is determined from the condition of coincidence of the MLO width $\Gamma_\gamma\left(E_\gamma=E_r\right)$ at GDR energy with the GDR width $\Gamma_r$ in cold nuclei: $k_r=(\Gamma_r-C_{coll}E_r^2)\tau_w/\hbar$. Values for $k_0=0.3$, $n_s=1$ and $F=1$ were found by comparison with experimental $\gamma$ -decay strengths.

It should be mentioned that the zero energy limit of the γ-decay RSF within the microcanonical description of excited nuclear states (3), (5) is determined by the value of the relaxation function $\Phi''_{X\lambda}(\omega)\equiv\chi''_{X\lambda}(\omega)/\omega$ at zero frequency and for E1 transitions, it is equal to:

$$\overleftarrow{f}_{E1}(E_\gamma=0)=\frac{4\pi}{9}\frac{e^2}{(\hbar c)^3}\cdot T\cdot\lim_{\omega\to+0}\left\{\frac{S_{E1}(\omega)}{\hbar\omega}\right\}\equiv-\frac{4}{9\hbar}\frac{e^2}{(\hbar c)^3}\cdot T\cdot\Phi''_{E1}(\omega=0)\ . \quad (33)$$

Note that the zero frequency limit of the relaxation function also determines the friction coefficient of the corresponding mode of collective motion [39,40].

In accordance with (22), (33), we have for the MLO model:

$$\overleftarrow{f}_{MLO}(E_\gamma=0)=8.674\cdot10^{-8}\cdot\sigma_r\Gamma_r\frac{\Gamma_\gamma(E_\gamma=0)}{E_r^3}\frac{T}{E_r}\quad(MeV^{-3})\,. \quad (34)$$

This expression differs from those obtained within the KMF, EGLO and GFL models (Eq.(20)) by the factor $(\Gamma_\gamma(E_\gamma=0)/\Gamma_\alpha(E_\gamma=0))T/(E_rK_\alpha)$.

## 3. Calculations and discussion

In this section, we test the various expressions of the dipole RSF for the photoabsorption and $\gamma$ -emission processes, as described in the previous section.

Figure 1 presents the relative deviations Δ of the photoabsorption cross sections calculated within different methods with respect to the MLO1 prediction:

$$\Delta=\left[\frac{1}{N_{\max}}\sum_{i=1}^{N_{\max}}\left[\frac{\sigma(A_i,Model)-\sigma(A_i,MLO1)}{\sigma(A_i,MLO1)}\right]^2\right]^{1/2}, \quad (35)$$

where $N_{\max}$ is the number of atomic nuclei with mass number $A_i$ involved in the calculation.

The relative deviations are compared at three γ-ray energies: 1) close to neutron separation energy $\left(E_\gamma=7\ MeV\right)$; 2) at the GDR energy $\left(E_\gamma=E_r\right)$ and 3) at an energy higher than the GDR energy by the width $\left(E_\gamma=E_r+\Gamma_r\right)$. The calculations were performed for atomic nuclei both on the beta-stability line (panels (a)) and with allowance of the isotopes out of this range (panels (b)). In the last case the calculations were done for the 3317 nuclei with $8\leq Z\leq 84$ lying between the proton and neutron drip-lines and listed on the website of Ref.[9]. The approximation of axially deformed nuclei was used for the RSF calculations in deformed nuclei [3].

The calculations within the HFB-QRPA approach are described in Refs.[9,10]. They are performed with the use of the Skyrme forces and an additional folding procedure to take collisional damping into account.

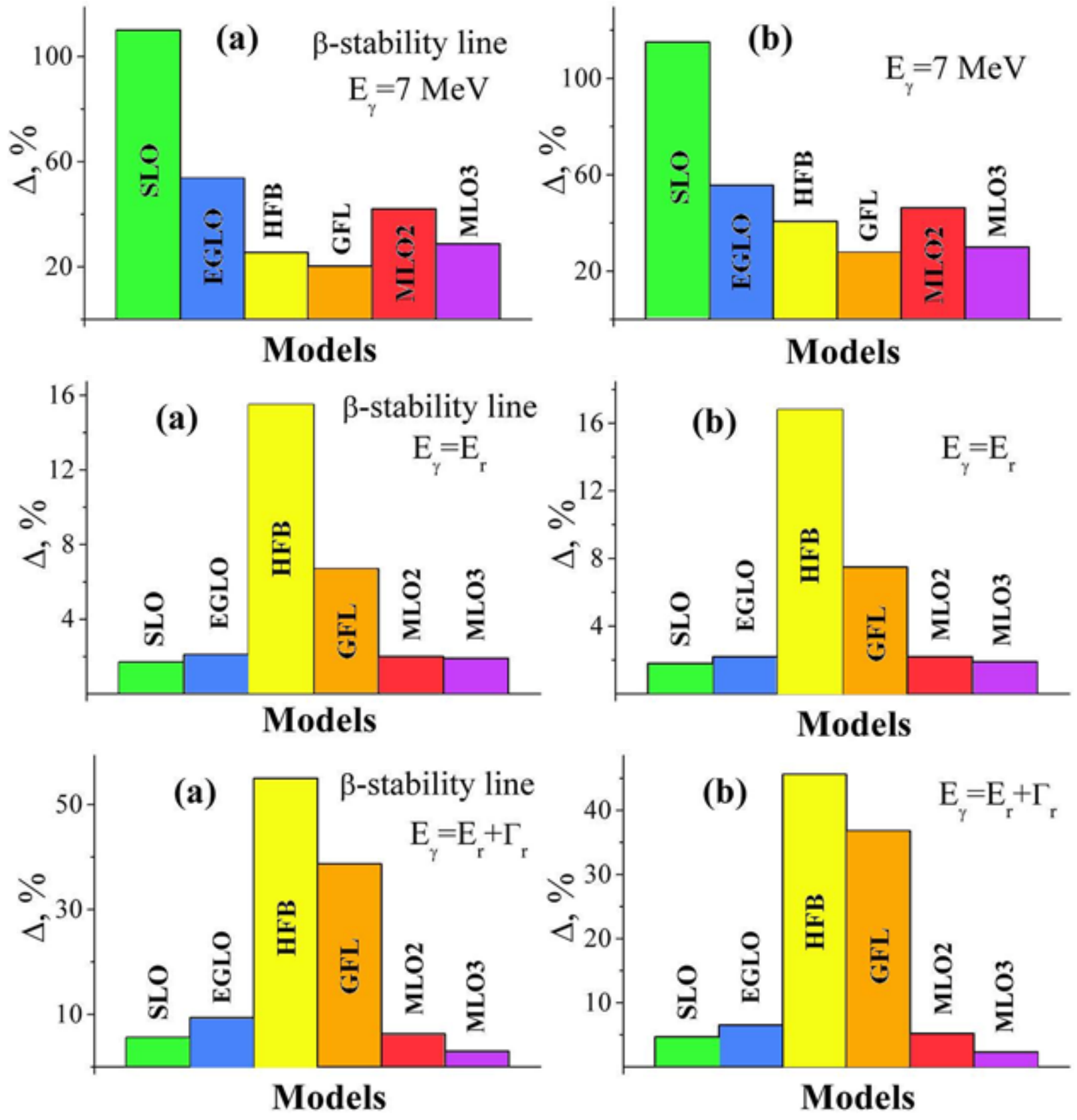

Fig. 1. Relative deviation of RSF calculated within different models from the MLO1 predictions for nuclei on $\beta$-stability line (a) and for isotopes between neutron and proton drip-lines (b) at $E_\gamma = 7$ MeV, $E_\gamma = E_r$ and $E_\gamma = E_r + \Gamma_r$.

The GDR parameters in the calculations of the EGLO, GFL and MLO models are taken from the experimental compilation RIPL[2,3] based on Refs.[17-18] (if more than one set of the parameters is given for an isotope, we adopt the values of the first set). These parameters for spherical nuclei are also presented in the Table B1 of the Appendix B in the line "SLO". In absence of experimental data, the GDR parameters are estimated from the following systematics [3,17]:

$$E_r = 31.2 / A^{1/3} + 20.6 / A^{1/6} \text{ (MeV)}, \ \Gamma_r = 0.026 E_r^{1.91} \text{ (MeV)}. \qquad (36)$$

The GDR peak cross sectiont is obtained from the relations

$$\frac{\pi}{2}\sigma_r \Gamma_r = 1.2 \cdot \sigma_{TRK}, \quad \sigma_{TRK} = 60 \cdot \frac{NZ}{A} \text{ (mb)}, \qquad (37)$$

where $\sigma_{TRK}$ is the classical dipole Thomas-Reiche-Kuhn sum rule and the factor 1.2 takes into account contribution of the velocity-dependent and exchange components of the nuclear forces.

The shape parameters of the SMLO model are obtained by fitting the theoretical calculations for photoabsorption cross sections to the experimental data from the EXFOR library (http://www-nds.iaea.org/exfor/). The ready-to-use tables (B1 and B2) of these parameters are presented in the Appendix B. It can be seen from the tables that the SMLO parameters are in rather close agreement with the SLO parameters with deviations of about $\sim 10-15\%$. Differences in GDR parameters for the SLO and SMLO models demonstrate the impact of the energy dependence of the width $\Gamma_\gamma(E_\gamma)$. Since this energy-dependence of $\Gamma_\gamma(E_\gamma)$ is not quite known, these devations can be considered as the current physical uncertainty remaining when extracting the GDR parameters ($\Gamma_r$, $E_r$ and $\sigma_r$) from a fit to experimental data.

In Fig.1, the photoabsorption RSF calculated within different closed-form models are compared. It can be seen that the dfferent predictions are in rather close agreement in a range of $\gamma$-ray energies around the GDR peak. However in the low-energy region, the MLO and EGLO predictions differ from the SLO ones. In particular, at $E_\gamma = 7\ MeV$ the RSF within the SLO approach is about twice larger with respect to the one obtained within the MLO and EGLO models (excluding the 150 ≤A≤180 mass range for the EGLO model). It should be pointed out that the cross section peak given by the HFB-QRPA model are, as a rule, can differ by 0.5 to 1 MeV from the GDR energy. It should be recalled here that the HFB-QRPA model significantly differs from all other models in the sense that the energy centroid is, in contrast to phenomenological models, predicted and not deduced from experiemtnal data or systematics. Differences of the order of 0.5 to 1 MeV correspond to the level of accuracy with which the GDR energy can be estimated by a mean field model like HFB+QRPA. It should be also mentioned that a collisional component of the GFL damping width can be negative in some deformed nuclei.

Figures 2, 3 compare the photoabsorption cross sections for $^{40}$Ca and $^{208}$Pb.

Different variants of MLO model give similar trend for photoabsorption cross sections. Therefore, only the MLO1 calculations are shown in the figures.

The following values of the parameters of the MSA model were used in the calculations: the Fermi energy 36.9 *MeV*; effective nucleon mass $m^* = 0.9$; the Landau parameters $F_0' = 0.929$, $F_0 = -0.227$ for the isovector and isoscalar components of the nucleon interaction in symmetric nuclear matter; the magnitude of symmetry surface energy is equal to $Q = 70\ MeV$. The relaxation time approximation was adopted for the collision integral with a constant value of the collective state relaxation time corresponding to the systematics of the GDR widths. The RSF within the semi-classical MSA method [11,12] includes two contributions, namely, a volume component that is related to the shift of proton and neutron fluids in the nuclear interior and a surface component due to vibrations of mutually non-penetrating neutron and proton spheres.

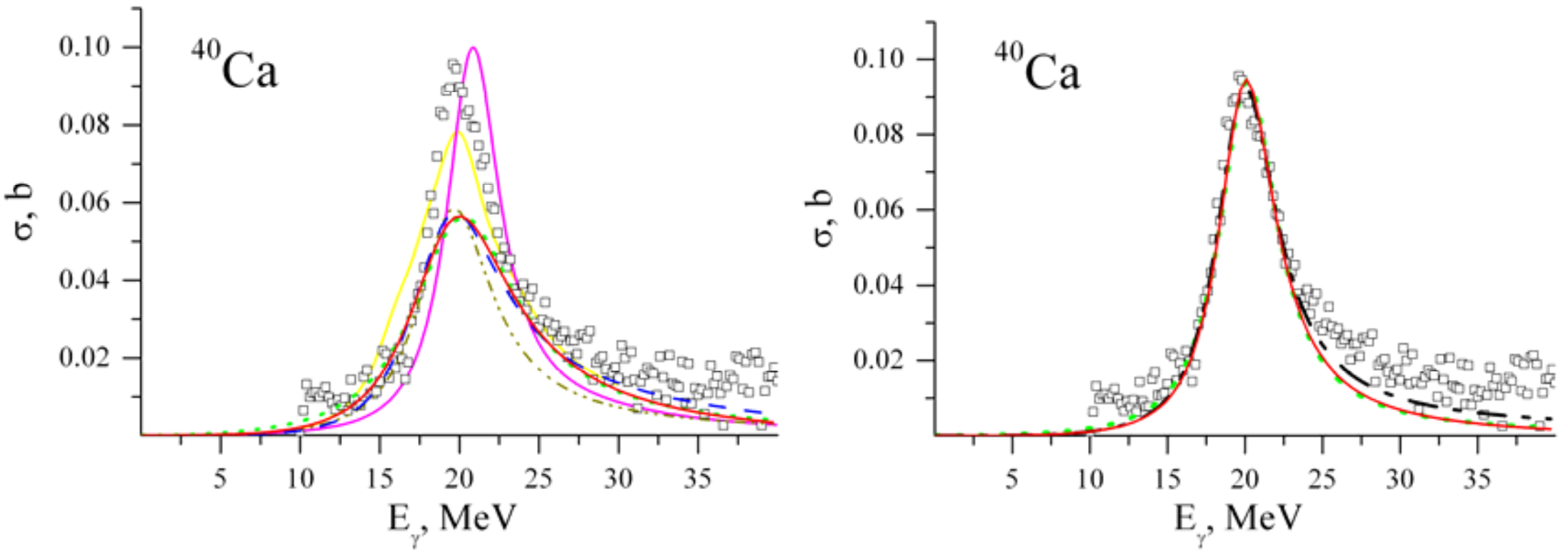

*a* *b*

**Fig. 2.** Comparison of the photoabsorption cross sections on $^{40}$Ca. Panels: *a* shows calculations with GDR parameters from systematics (36),(37); *b*- calculations with GDR parameters obtained from fitting the data from [41] (see below Table B1). Curves: red solid line- MLO1; black dash-dot line SMLO, green dot line – SLO; blue dash line - EGLO, grey dash-dot-dot line - GFL, yellow solid line- HFB-QRPA, magenta solid line- MSA. Points are experimental data estimated from [41].

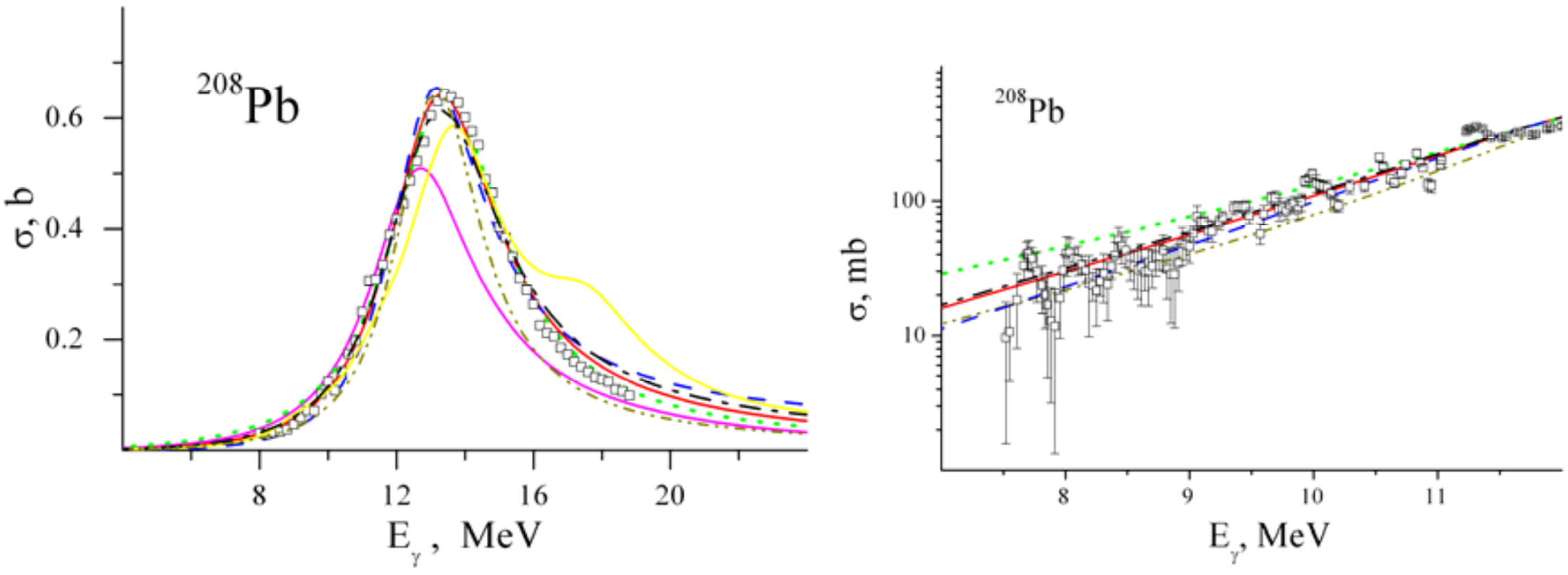

*a* *b*

**Fig. 3.** Comparison of the photoabsorption cross sections on $^{208}$Pb. Panel *b* shows the low-energy part of the cross sections. Experimental data are taken from [42] in panel *a* and from [43] in panel *b*. The SLO parameters are taken from he RIPL library [2,3]. Notations are the same as in Fig.2.

Calculations of the photoabsorption cross sections within closed-form models with GDR parameters from systematics (36),(37) ) ( $E_r = 20.26\,\text{MeV}$, $\Gamma_r = 8.14\,\text{MeV}$, $\sigma_r = 56.3\,\text{mb}$ ) presented on panel *a* of Fig.2. Panel *b* of Fig.2 demonstrates calculations with the GDR parameters obtained from fitting the data from [41] and they are presented in the Table B1 of the Appendix B.

It can be seen that the RSF calculations are rather sensitive to set of the GDR parameters. We see that standard GDR systematics (36), (37) is poor to describe photoabsorption cross sections in the nuclei (like $^{40}$Ca) with a major contribution of the photo-charged-particle reaction cross sections to $\sigma(\gamma, abs)$. On average, both the MSA and HFB-QRPA calculations agree with the experimental data and gives better predictions in this range of the atomic nuclei without additional corrections of global parameters. The results of the photoabsorption RSF calculations within the MLO1 and SMLO are in a close agreement. In line with the previous investigations [2,3,27,44,45], the calculations within renewed closed-form models (EGLO, GFL, MLO and SMLO) at γ-ray energies close to neutron separation energy are, as a rule, in better accordance with experimental data than calculations within SLO model with GDR parameters (standard SLO). However, as can be seen from Table B2 (see Appendix B), photoabsorption data in a rather wide energy range up including the neutron separation energy, can also be fitted by the SLO model, but, as a rule, with a smaller width as compared to the GDR value.

Fig.4 demostrates the γ-ray energy dependence of the E1 strength function and photoabsorption cross section for $^{144}$Nd. The experimental data are taken from Fig.7 of the Ref. [19] for γ-decay RSF at excitation energy near the neutron separation energy and from the Ref.[46] for photoabsorption cross section. Experimental data for electric dipole RSF of the γ-decay process were obtained in Ref.[19] under assumption that the ratio of the electric to the magnetic RSF is equal to unity.

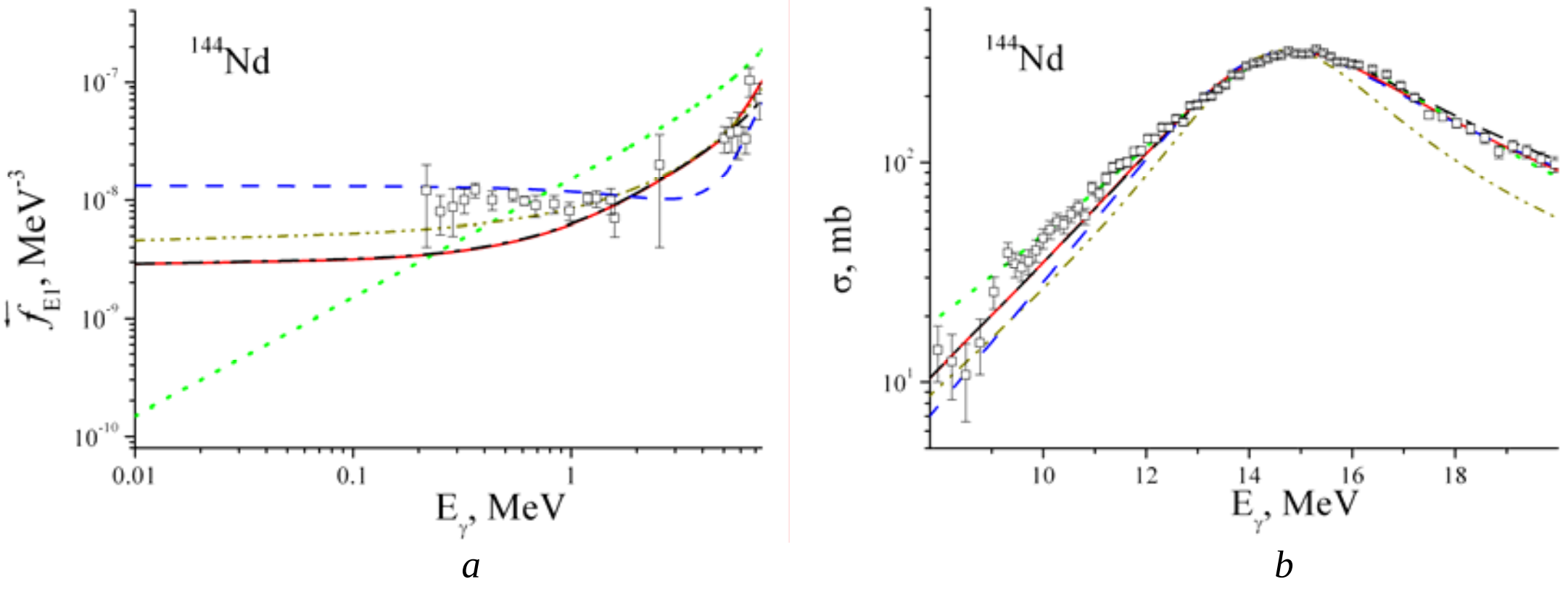

**Fig. 4.** The E1 γ-decay strength function at $U = S_n = 7.8$ MeV (left panel) and photoabsorption cross section (right panel) for $^{144}$Nd. The experimental data are taken from Fig.7 of the Ref. [19] in panel *a* and from the Ref.[46] in panel *b*. Notations are the same as in Fig.2.

In these calculations, GDR parameters are derived, as previously, by a fit to experimental photoabsorption data, as given in Table B1. The temperatures are determined by the equation

$$U - U_s = aT^2, \tag{38}$$

where $U$ is the corresponding excitation energy, $U_s$ the energy shift of the back-shifted Fermi-gas model and $a$ the energy-dependent level density parameter. Values for the level density parameters and energy shifts are taken from the "beijing_bs1.dat"- file of the RIPL1 compilation [2] (assuming a rigid-body moment of inertia) or from global systematics [47], when no experimental data are available.

The EGLO, GFL, MLO and SMLO results of the calculations for γ–decay are all characterized by a non-zero limit and a temperature dependence at low γ-ray energies. All these models describe the experimental data much better than the SLO model, which predicts a vanishing strength function at zero $\gamma$-ray energy.

Fig.5 shows dipole $\gamma$-decay strength functions plotted against mass number for 7 nuclei ($^{91}$Zr, $^{144}$Nd, $^{146}$Nd, $^{148}$Sm, $^{150}$Sm, $^{207}$Pb and $^{208}$Pb) considered as spherical ones. The results were calculated for γ-ray energies that approximately correspond to the mean energy of E1 transitions in the file "gamma-strength-exp.dat" from the RIPL2 library [3]. Table 1 shows least-square deviations (Eq.(B9) with $N_{par} = 0$) of the calculated RSF with respect to experimental data.

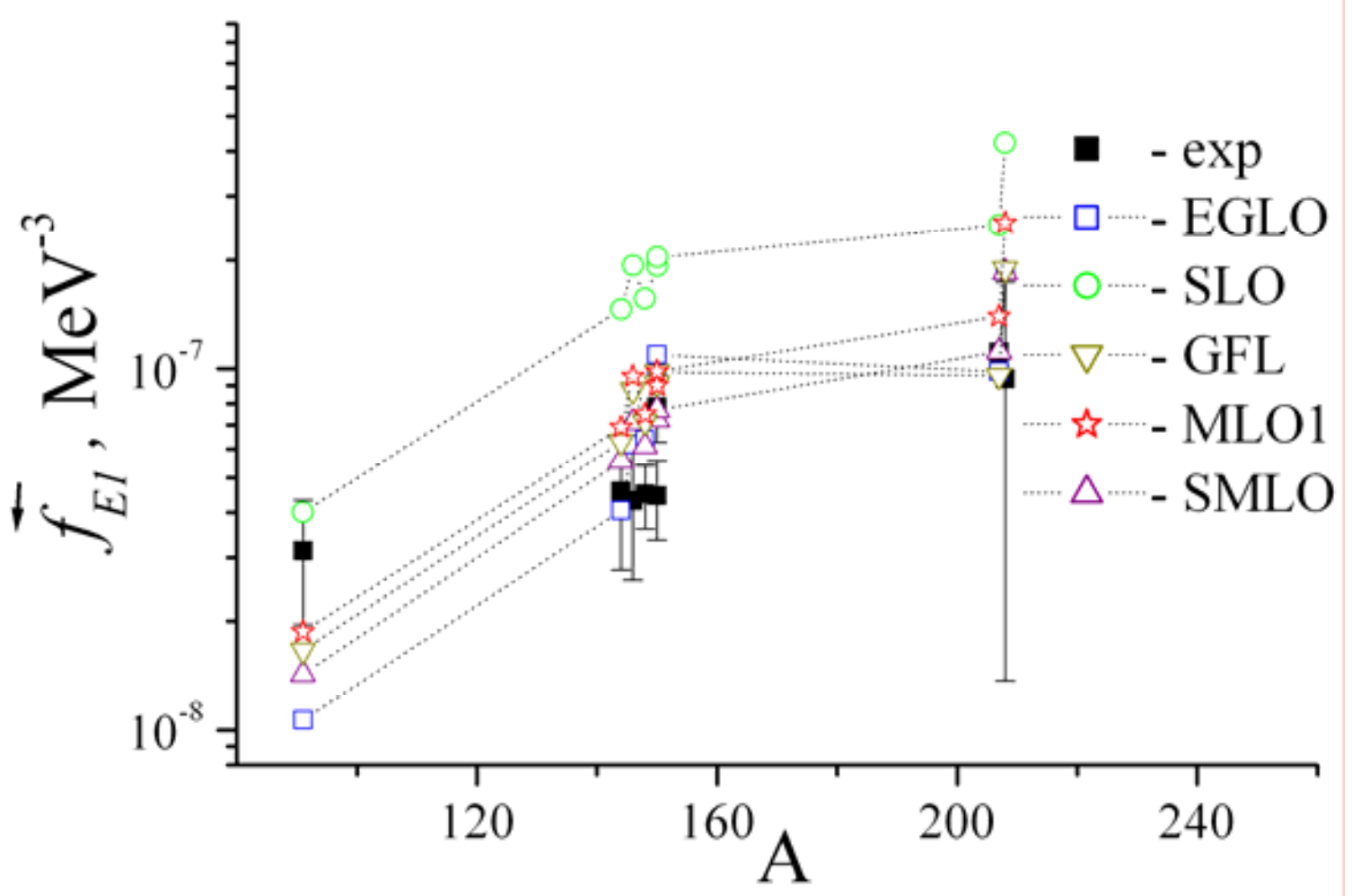

**Fig. 5.** The E1 γ-decay strength function versus mass number for 7 nuclei ($^{91}$Zr, $^{144}$Nd, $^{146}$Nd, $^{148}$Sm, $^{150}$Sm, $^{207}$Pb and $^{208}$Pb) considered as spherical ones; $U = S_n$, $E_\gamma = 0.8U$ . The experimental data are taken from "gamma-strength-exp.dat" file of the RIPL2[3].

**Table 1.** The $\chi^2$ deviations of the theoretical γ-decay strength functions from experimental data for nuclei $^{91}$Zr, $^{144}$Nd, $^{146}$Nd, $^{148}$Sm, $^{150}$Sm, $^{207}$Pb and $^{208}$Pb

| Model | SLO | EGLO | GFL | MLO1 | SMLO |
|---|---|---|---|---|---|
| $\chi^2$ | 105.0 | 5.0 | 5.27 | 7.55 | 2.00 |

Figure 5 and Table 1 show, in agreement with the previous investigations [2,3,27,48], that all considered models with asymmetric shape of the RSF (EGLO, GFL, MLO1, SMLO) describe the experimental γ-decay data with $E_\gamma \sim S_n$ better than the standard SLO model based on GDR parameters.

The overall comparison of the calculations within different simple models and experimental data shows that the EGLO and MLO (SMLO) approaches with asymmetric shape of the RSF provide a unified and rather reliable simple method to estimate the dipole RSF both for γ-decay and for photoabsorption over a relatively wide energy interval ranging from zero to slightly above the GDR peak, at least, when GDR parameters are known or GDR systematics can be safely applied to. Otherwise, the HFB-QRPA and semi-classical MSA seem to be more adequate to describe the dipole photoabsorption RSF in spherical nuclei of medium mass.

It can be noted that different variants of the MLO (SMLO) approach are based on general relations between the RSF and the nuclear response function. Therefore they can potentially lead to more reliable predictions among simple models. However, the energy dependence of the width $\Gamma_\gamma(E_\gamma)$ is governed by complex mechanisms of nuclear dissipation and is still an open problem.

Reliable experimental information is needed to better determine the temperature and energy dependence of the RSF, so that the contributions of the different mechanisms responsible for the damping of the collective states can be further investigated. This should help us to discriminate between the various closed-form models describing the dipole RSF.

## 4. Acknowledgment

VP is very thankful to Profs. A.V.Ignatyuk, P.Oblozinsky, J.Kopecky, M.Herman, R.Capote Noy, M.G. Urin, V.I.Abrosimov, A.G.Magner and F.A.Ivanyuk for valuable discussions and Prof. F. Becvar for support and the kind hospitality at the Prague Workshop.

This work is supported in part by the IAEA(Vienna) under IAEA Research Contract No.12492.

## Appendix A. MLO strength function as simplified version of semi-classical SRPA

In order to obtain a simple expression for the linear response function $\chi_{EL}(\omega)$, we will use a semi-classical approach [32] based on the Landau-Vlasov kinetic equation for the nucleon phase-space distribution function $f(\vec{p},\vec{r},t)$ completed by collision term $J_c(\vec{p},\vec{r},t)$:

$$\frac{\partial f}{\partial t}+\frac{\vec{p}}{m}\cdot\vec{\nabla}_r f-\vec{\nabla}_r\left[V(\vec{r},t)+V^L_{ext}(t)\right]\cdot\vec{\nabla}_p f=J_c. \tag{A1}$$

with $V$ for the self-consistent mean field. The collision integral is taken in non-Markovian form with allowance for retardation effects, that is, with dependence of its magnitude at time $t$ on the distribution function at the previous times $t'$ ([33,34,49,50 and Refs. therein]

$$J_c(\vec{p},\vec{r},t)=\int_{-\infty}^{t} dt' A(t-t')\, f\left(\vec{r},\vec{p},t'\right), \tag{A2}$$

where a kernel $A(t-t')$ is responsible for the memory effects. The Fourier transform $J_c(\vec{p},\vec{r},\omega)$ of this collision integral has the form of a relaxation time approximation

$$J_c(\vec{p},\vec{r},\omega) = -\frac{f\left(\vec{r},\vec{p},\omega\right)}{\tau_c(\omega)} \tag{A3}$$

with frequency dependent relaxation time $\tau_c(\omega)$ which shape is determined by kernel $A$.

Following the mathematical treatment of the Ref.[32] but with frequency dependent relaxation time, we find expression for linear response function $\chi_{EL}(\omega)$ of the same form as in [32]:

$$\chi_{EL}(\omega) = \frac{\bar{\chi}_L(\omega)}{1-k_L\bar{\chi}_L(\omega)}. \tag{A4}$$

The strength function $S_{EL}(\omega)$, Eq.(5), for response function of this form can be presented as

$$S_{EL}(\omega) = -\frac{k_L^{-1}}{\pi} Im\frac{1}{1-k\bar{\chi}_L} = -\frac{k_L^{-2}}{\pi}\frac{\bar{\chi}_L''}{(\bar{\chi}_L'-k_L^{-1})^2+((\bar{\chi}_L'')^2}, \tag{A5}$$

where $\bar{\chi}_L' \equiv Re\bar{\chi}_L$ and $\bar{\chi}_L'' \equiv Im\bar{\chi}_L$.

The quantity $k_L$ in Eq.(A4) is the coupling constant of the coherent separable interaction $V_L(r,r')$ between two particles with the same symmetry as the external field acting on the nucleus,

$$V_L(r,r') = k_L Q_L(r) Q_L(r'). \tag{A6}$$

In the random phase and self-consistent Hartree-Fock approximations this type of interaction redistributes the one-particle one-hole (*1p-1h*) excitations to form the collective vibrational states[51,52].

The function $\bar{\chi}_L(\omega)$ in Eqs.(A4), (A5) is the so-called uncorrelated (intrinsic) response function. It describes the response of the nuclear system without coherent interaction (A6) but in presence of residual incoherent two-body interaction $V_{res}$ that couples the 1p1h excitations to more complex states lying at the same excitation energy. In the kinetic theory[32], the residual interaction is simulated by the collision integral. An additional folding procedure is used in the HFB-QRPA [9,10] to take into account this collisional damping.

The semi-classical intrinsic response function has the following form (in approximation $N = Z = A/2$):

$$\bar{\chi}_L(\omega) = -\frac{4}{3\pi\hbar^3}\int_0^\infty d\varepsilon\frac{d\bar{f}}{d\varepsilon}\int_0^{l_m(\varepsilon)} dl\, l\, T(\varepsilon,l) P_L(\varepsilon,l), \tag{A7}$$

and it is determined by a motion of the nucleons on single-particle orbits with angular momentum $l$ and energy $\varepsilon$; $l_m(\varepsilon)$ is maximal angular momentum and

$$P_L(\varepsilon,l) = \sum_{N=-L}^{L} Y_{LN}(\pi/2,\pi/2)^2 \sum_{n=-\infty}^{\infty} \bar{Q}_L(nN;\varepsilon,l)^2 \frac{\bar{\omega}_{nN}(\varepsilon,l)}{\omega-\bar{\omega}_{nN}(\varepsilon,l)}. \tag{A8}$$

Here, $Y_{LN}$ is the spherical function; $\bar{Q}_L(nN;\varepsilon,l)$ semi-classical radial matrix element of the form:

$$\bar{Q}_L(nN;\varepsilon,l)=\frac{2}{T(\varepsilon,l)}\int_{r_1}^{r_2}dr\frac{Q_L(r)}{v(r,\varepsilon,l)}\cos[\omega_{nN}(\varepsilon,l)\tau(r,\varepsilon,l)-N\gamma(r,\varepsilon,l)]\,, \qquad (A9)$$

where in accordance with corrections from Refs.[11,12,53] the dependence on damping is absent.

The $\varpi_{nN}(\varepsilon,l)$ is a complex frequency:

$$\varpi_{nN}(\varepsilon,l)=\omega_{nN}(\varepsilon,l)-i\eta,\;\; \eta=\frac{1}{\tau_c(\omega=E_\gamma/\hbar)} \qquad (A10)$$

with relaxation time $\tau_c$ that can be dependent on the γ-ray energy due to presence of the retardation effects in the collision integral; $\omega_{nN}(\varepsilon,l)$ is the independent-particle frequency:

$$\omega_{nN}(\varepsilon,l)=n\omega_0(\varepsilon,\lambda)+N\omega_v(\varepsilon,\lambda),\; \omega_0(\varepsilon,\lambda)=\frac{2\pi}{T(\varepsilon,l)}\;,\; \omega_v(\varepsilon,\lambda)=\omega_0(\varepsilon,\lambda)\Gamma(\varepsilon,l)\,. \qquad (A11)$$

The quantities $\tau(r,\varepsilon,l)$ and $\gamma(r,\varepsilon,l)$ in (A6) are respectively the time elapsed and the angle spanned to reach position $r$ on the orbit $(\varepsilon,l)$ for a nucleon with the radial velocity in the static self-consistent potential of the mean field for independent particles. The quantities $T(\varepsilon,l)$ and $\Gamma(\varepsilon,l)$ in (A11) are the period of radial and angular motions, respectively.

It can be seen that following conditions of symmetry are fulfilled for frequency $\omega_{nN}(\varepsilon,l)$, matrix element $\bar{Q}_L(nN;\varepsilon,l)$ and the spherical function

$$\omega_{-n-N}=-\omega_{nN}\,,\;\; \bar{Q}_L(nN)=\bar{Q}_L(-n-N)\,,\;\; Y_{L-N}(\pi/2,\pi/2)=(-1)^N\cdot Y_{LN}(\pi/2,\pi/2)\;. \qquad (A12)$$

With the use of these relationships and with combining terms with different sign of $n$ in (A8), the quantity $P_L$ can be written in resonance form with a Lorentzian shape for the individual resonances:

$$P_L(\varepsilon,l)=P_L^{(0)}(\varepsilon,l)+\sum_{N>0}^{L}|Y_{LN}|^2\sum_{n=-\infty}^{\infty}|\bar{Q}_L(nN;\varepsilon,l)|^2\frac{Z_n(N)}{(\omega^2-\Omega_n^2(N))^2+4\eta^2\omega^2} \qquad (A13)$$

where $\Omega_n^2(N)\equiv|\varpi_{nN}(\varepsilon,l)|^2=\omega_{nN}^2+\eta^2$,

$$Z_n(N)=[\Omega_n^2(N)(\omega^2-\Omega_n^2(N))-2\eta^2\omega^2]-i2\eta\omega[\omega^2+\Omega_n^2(N)]\,, \qquad (A14)$$

and

$$P_L^{(0)}(\varepsilon,l)=-i|Y_{L0}|^2|\bar{Q}_L(00;\varepsilon,l)|^2\frac{\eta}{\omega+i\eta}+|Y_{L0}|^2\sum_{n=1}^{\infty}|\bar{Q}_L(n0;\varepsilon,l)|^2\frac{Z_n(0)}{(\omega^2-\Omega_n^2(0))^2+4\eta^2\omega^2}\,. \qquad (A15)$$

The spherical functions $Y_{LN}\equiv Y_{LN}(\pi/2,\pi/2)$ vanish unless $N$ has the same parity as $L$. Therefore, the summing over $N$ in Eq. (A13) involves only terms with either odd or even $N$ and the function $P_L^{(0)}$ is not equal to zero only for even $L$, i.e., $P_{L=2j+1}^{(0)}(\varepsilon,l)=0,\;\; j\geq 0$.

Using these equations, the imaginary and real parts of the expressions (A7), (A8) for dipole intrinsic response function can be present as:

$$\bar{\chi}'_{L=1}=\langle\langle\bar{P}'(\omega,\eta)\rangle\rangle\,,\;\; \bar{P}'(\omega,\eta)\equiv 2\sum_{n=-\infty}^{\infty}|\bar{Q}(n)|^2\frac{\Omega_n^2(\omega^2-\Omega_n^2)-2\eta^2\omega^2}{(\omega^2-\Omega_n^2)^2+4\eta^2\omega^2}\,, \qquad (A16)$$

$$\bar{\chi}''_{L=1} = \ll \bar{P}''(\omega,\eta) \gg, \quad \bar{P}''(\omega,\eta) = -\sum_{n=-\infty}^{\infty} |\bar{Q}(n)|^2 \frac{2\eta\omega(\omega^2+\Omega_n^2)}{(\omega^2-\Omega_n^2)^2+4\eta^2\omega^2}. \qquad (A17)$$

Here, the symbol $\ll ... \gg$ denotes the integrals of the following structure

$$\ll ... \gg = -\frac{1}{2\pi^2\hbar^3}\int_0^{\infty} d\varepsilon \frac{d\bar{f}}{d\varepsilon}\int_0^{l_m(\varepsilon)} dl\, l\, T\cdot(..) \qquad (A18)$$

over the single-particle energy and angular momentum and

$$\Omega_n^2 \equiv \Omega_n^2(N=1) = \omega(n)^2+\eta^2, \quad \omega(n)=\omega_{nN=1}, \quad \bar{Q}(n) \equiv \bar{Q}_{L=1}(nN=1;\varepsilon,l). \quad (A19)$$

The semiclassical dipole strength function is calculated using Eq.(A5) with (A16)-(A19).

In order to obtain simple closed-form expression for the dipole strength function near a collective resonance, we assume that 1) all particle- hole energies are degenerated at the frequency $\omega_0 = E_0/\hbar$ ($\omega(n)=\omega_0$) and 2) collective energy and energy weighted sum rule are weakly affected by the damping.

With these approximations and according to the RPA approach[51,52], a magnitude of GDR frequency $\omega_r = E_r/\hbar$ is found from the following dispersion relation

$$k_1^{-1} = \bar{\chi}_1'(\omega) = \ll \bar{P}'(\omega,\eta\to+0) \gg =$$

$$= \ll \sum_{n=-\infty}^{\infty} \frac{2\omega^2(n)\bar{Q}^2(n)}{\omega^2-\omega^2(n)} \gg = \frac{2\omega_0^2}{\omega^2-\omega_0^2} \ll \sum_{n=-\infty}^{\infty}\bar{Q}^2(n) \gg, \qquad (A20)$$

which corresponds to the singularity condition of the response function (A5) in the weak damping case.

The integrated sum of squared matrix element is proportional to the energy weighted sum rule (EWSR) of non-interacting particles $\bar{S}_{EWSR}$ (Thomas- Reiche- Kuhn (TRK) sum rule for strength function):

$$\ll \sum_{n=-\infty}^{\infty}\bar{Q}^2(n) \gg = \bar{S}_{EWSR}/(\hbar\omega_0)^2, \qquad (A21)$$

where

$$\bar{S}_{EWSR} \equiv \hbar^2\int_0^{\infty} d\omega\left[-\frac{1}{\pi}\bar{\chi}''_{L=1}(\omega)\right] = \hbar^2 \ll \sum_{n=-\infty}^{\infty}\omega^2(n)\bar{Q}^2(n) \gg. \qquad (A22)$$

Resulting expression for collective energy has the standard RPA form for system with degenerate particle-hole states:

$$E_r^2 = E_0^2 + 2k_1\bar{S}_{EWSR}. \qquad (A23)$$

Then the first term of denominator in Eq. (A5) in the vicinity of a resonance energy takes the form

$$D(\omega) \equiv k_1\bar{\chi}'_{L=1} - 1 \approx k_1 \ll Re\bar{P}(\omega,\eta\to+0) \gg -1 = \frac{E_r^2-E_\gamma^2}{E_\gamma^2-E_0^2} \approx \frac{E_r^2-E_\gamma^2}{2k_1\cdot\bar{S}_{EWSR}}. \qquad (A24)$$

With the use of this equation, the dipole strength function can finally be represented by the following Lorentzian form:

$$S_{EL=1}(E_\gamma) = \frac{2}{\pi}\bar{S}_{EWSR}\frac{E_\gamma \Gamma(E_\gamma)}{(E_\gamma^2 - E_r^2)^2 + (\Gamma(E_\gamma)E_\gamma)^2}. \qquad \text{(A25)}$$

Here, the $\Gamma \equiv \Gamma(E_\gamma)$ is the energy dependent width:

$$\Gamma(E_\gamma) = \frac{2k_1^2 \bar{S}_{EWSR}}{\hbar} << \sum_{n=-\infty}^{\infty} |\bar{Q}(n)|^2 \frac{2\eta\omega(\omega^2 + \Omega_n^2)}{(\omega^2 - \Omega_n^2)^2 + 4\eta^2\omega^2} >> \approx \frac{\tilde{\alpha} 2\hbar\eta(E_\gamma^2 + E_0^2 + (\hbar\eta)^2)}{(E_\gamma^2 - E_0^2 - (\hbar\eta)^2)^2 + 4(\hbar\eta E_\gamma)^2}$$

or

$$\Gamma(E_\gamma) = \hbar\frac{\beta(E_\gamma;\tau_c)}{\tau_c(E_\gamma)}, \qquad \text{(A26)}$$

where

$$\beta(E_\gamma;\tau_c) = \frac{2\tilde{\alpha}(E_r^2 + E_0^2)}{(E_r^2 - E_0^2)^2 + 4(E_\gamma\hbar/\tau_c(E_\gamma))^2}, \quad \tilde{\alpha} = \frac{2k_1^2 \bar{S}_{EWSR}}{E_0^2} = \frac{E_0^2}{2}\left(1 - \frac{E_r^2}{E_0^2}\right)^2. \qquad \text{(A27)}$$

The quantity $\Gamma(E_\gamma)$ at the energy $E_\gamma = E_r$ can be considered as the GDR width. It has form of the expression for width obtained in Ref. [54].

The MLO expressions (24), (27) for the dipole radiative strength functions are obtained from Eqs.(A25), (A26) with the magnitude of the energy weighted sum rule (EWSR) $\bar{S}_{EWSR}$ that corresponds to the expression $(\pi/2)\sigma_r\Gamma_r$ giving the area under the Lorentz curve in the case of a constant resonance width or a strong energy-concentrated GDR.

## Appendix B. Tables of GDR parameters in spherical nuclei

In this appendix, the GDR parameters in spherical nuclei are tabulated. The parameters are obtained by a fit of the theoretical photoabsorption cross sections within the SMLO and SLO models to the experimental data. The calculations were performed for 49 spherical (with quadrupole deformation parameters $|\beta_2| < 0.1$) as well as slightly deformed nuclei ($^{70,72}Ge$, $^{98,100}Mo$, $^{103}Rh$ $^{107}Ag$, $^{146}Nd$, $^{148,150}Sm$) when photoabsorption cross sections can be approximated by a single resonance curve[17,18].

In line with Refs.[17,18,55], dipole photoabsorption cross-section $\sigma_{E1}$ is taken to be equal to the total photoabsorption cross-section $\sigma(\gamma,abs)$ which is approximated (for all considered nuclei except $^{40}Ca$ ) by the total photoneutron cross section $\sigma(\gamma,sn)$

$$\sigma_{E1} \cong \sigma(\gamma,abs) \cong \sigma(\gamma,sn) = \sigma(\gamma,1nx) + \sigma(\gamma,2nx) + \sigma(\gamma,3nx) + ... + \sigma(\gamma,F), \qquad \text{(B1)}$$

where $\sigma(\gamma,F)$ is the total photofission cross section and $\sigma(\gamma,Nnx)$ the a sum of all cross sections leading to the ejection of $N$ neutrons, i.e.

$$\sigma(\gamma,Nnx) = \sigma(\gamma,Nn) + \sigma(\gamma,Nn\,p) + \sigma(\gamma,Nn\alpha) + ... \; . \qquad \text{(B2)}$$

The relationship (B1) is realized with a good accuracy due to small contributions of the photo-charged-particle reaction cross sections to $\sigma(\gamma,abs)$ for all considered nuclei except $^{40}Ca$ at γ-ray energies up to sligthly above the peak energy which is chosen as an upper limit for the fitting procedure. Experimental data on total photoneutron cross sections $\sigma(\gamma,sn)$ are taken from the international nuclear data library EXFOR (http://www-nds.iaea.org/exfor/). The photoabsorption cross section for $^{40}Ca$ is taken from Ref.[41] and it includes contribution of $(\gamma,p)$ reaction.

For the nuclei $^{70,72}Ge$, $^{80,82}Se$, $^{93}Nb$, $^{94,100}Mo$, $^{133}Cs$, $^{124,126,128,130}Te$, $^{140,142}Ce$ and $^{146}Nd$, the EXFOR data base contains only information on photoneutron cross sections $\sigma(\gamma,1nx)$ and the inclusive photoneutron yield cross section $\sigma(\gamma,xn)$ which includes the multiplicity of neutrons emitted in each reaction event:

$$\sigma(\gamma,xn)=\sigma(\gamma,1nx)+2\sigma(\gamma,2nx)+3\sigma(\gamma,3nx)+...+\bar{\nu}(\gamma,F), \quad \text{(B3)}$$

with $\bar{\nu}$ for the average multiplicity of photofission neutron. For considered ranges of the γ-energy and nuclei, the cross section $\sigma(\gamma,xn)$ can be presented as

$$\sigma(\gamma,xn)=\sigma(\gamma,1nx)+2\sigma(\gamma,2nx). \quad \text{(B4)}$$

Therefore in order to obtain the total photoneutron cross section, we estimate of $\sigma(\gamma,2nx)$ by the use of Eq.(B4) in the following way

$$\sigma(\gamma,2nx)=[\sigma(\gamma,xn)-\sigma(\gamma,1nx)]/2, \quad \text{(B5)}$$

and then the $\sigma(\gamma,sn)$ is estimated by the expression

$$\sigma(\gamma,sn)=\sigma(\gamma,1nx)+\sigma(\gamma,2nx) \;. \quad \text{(B6)}$$

The errors $\Delta\sigma(\gamma,sn)$ of the $(\gamma,sn)$ cross sections are calculated by the expression

$$\Delta\sigma(\gamma,sn)=\sqrt{\Delta^2\sigma(\gamma,1nx)+\Delta^2\sigma(\gamma,2nx)}, \quad \text{(B7)}$$

where

$$\Delta\sigma(\gamma,2nx)=\frac{1}{2}\sqrt{\Delta^2\sigma(\gamma,xn)+\Delta^2\sigma(\gamma,1nx)} \quad \text{(B8)}$$

with the experimental errors $\Delta\sigma(\gamma,xn)$ and $\Delta\sigma(\gamma,1xn)$ from the EXFOR data base.

The adjustment is performed by the least square method minimizing the $\chi^2$ - value:

$$\chi^2=\frac{1}{N-N_{par}}\sum_{i=1}^{N}\left(\frac{\sigma_{theor}(E_{\gamma,i})-\sigma_{\exp}(E_{\gamma,i})}{\Delta\sigma_{\exp}(E_{\gamma,i})}\right)^2 \quad , \quad \text{(B9)}$$

where $\sigma_{theor}(E_{\gamma,i})$ is the theoretical cross sections at γ-ray energy $E_{\gamma,i}$, $\sigma_{\exp}(E_{\gamma,i})$ the experimental cross section, $\Delta\sigma_{\exp}(E_{\gamma,i})$ the corresponding statistical error, and $N$ the total number of data points. $N_{par}$ is the number of parameters deduced from the fit (the value $N_{par}$=3 is used corresponding to the SLO model). Note that estimated data are used for $^{40}Ca$ [41] , so in this case errors equal to ten percent of the cross section are adopted: $\Delta\sigma_{\exp}\left(E_{\gamma,i}\right) = 0.1\cdot\sigma_{\exp}\left(E_{\gamma,i}\right)$.

The best least square minimization was performed using the MINUIT package (http://wwwasdoc.web.cern.ch/wwwasdoc/minuit/minmain.html). The results are presented in Tables B1 and B2, where we indicate used model, the values *Z, A* with chemical symbol of isotope, the curve parameters and either SMLO parameter $a$ or quantity $\Gamma_r / E_r$ for SLO, $\chi^2$ values, γ-ray energy range of the fitting and the data reference.

The Table B1 lists the parameters of the SMLO model simultaneously with SLO parameters from photoneutron cross-section data. All SLO parameters (except $^{40}Ca$ set) were obtained in Ref.[18] and they correspond to the first line set for relevant isotope in the file "gdr-parameters-exp.dat" from the RIPL[2,3]. The quantities $\chi^2_{in}$ and $\chi^2_{out}$ are the magnitudes of $\chi^2$ calculated for fitting intervals indicated in this line and for supplementary interval. The supplementary intervals are given either in line "SLO" of the Table B1 or correspond to extended intervals from the Table B2. The shape parameter $\Gamma_r$ within SMLO model are calculated by the expression $\Gamma_r = aE_r$ with parameters $a$ from the fit.

The Table B2 gives the SLO parameters calculated by $\chi^2$ minimization in different $\gamma$-ray energy intervals. Symbol SLO* denotes set of the parameters obtained from our calculations in extended and small fitting intervals. In the second line, SLO parameters from the Table B1 are listed. They were obtained in the small fitting interval near peak energies in Ref.[18]. The other denotations are the same as in Table B1.

**Table B1.** GDR parameters within SMLO and SLO models from the $\chi^2$ minimization of the least square deviation of theoretical photoabsorption cross sections from experimental data

| Model | Z | A | EL | $E_r$ *MeV* | $\sigma_r$ *mb* | $\Gamma_r$ *MeV* | $a$ | $\chi^2_{in}$; $\chi^2_{out}$ | Fitting interval *MeV* | Data Ref. |
|---|---|---|---|---|---|---|---|---|---|---|
| SMLO | 20 | 40 | Ca | 20.17 | 92.61 | 4.85 | .24 | 2.88; 19.1 | 15 - 23 | [41] |
| SLO | 20 | 40 | Ca | 20.16 | 94.47 | 4.50 | .22 | 2.89; 12.4 | 15 - 23 | |
| | | | | | | | | | | |
| SMLO | 29 | 65 | Cu | 16.95 | 74.69 | 7.17 | .42 | .68; 3.71 | 14 - 20 | [56] |
| SLO | 29 | 65 | Cu | 16.70 | 75.20 | 6.89 | .41 | .78; 1.41 | 14 - 20 | |
| | | | | | | | | | | |
| SMLO | 32 | 70 | Ge | 17.02 | 90.16 | 7.50 | .44 | 1.39; 1.12 | 11.55 - 21 | [57] |
| SLO | 32 | 70 | Ge | 16.79 | 89.40 | 7.66 | .46 | 1.30; 3.26 | 13 - 21 | |
| | | | | | | | | | | |
| SMLO | 32 | 72 | Ge | 16.93 | 106.93 | 7.87 | .46 | .75; 1.03 | 10.47 - 21 | [57] |
| SLO | 32 | 72 | Ge | 16.67 | 107.00 | 7.68 | .46 | 1.25; 2.64 | 13 - 21 | |
| | | | | | | | | | | |
| SMLO | 34 | 80 | Se | 16.43 | 134.51 | 6.30 | .38 | .48; 9.83 | 13 - 17 | [57] |
| SLO | 34 | 80 | Se | 16.53 | 136.00 | 6.90 | .42 | 1.42; 2.90 | 13 - 17 | |
| | | | | | | | | | | |
| SMLO | 34 | 82 | Se | 16.77 | 146.95 | 6.16 | .37 | .77; 2.35 | 13 - 21 | [57] |
| SLO | 34 | 82 | Se | 16.65 | 148.00 | 5.91 | .35 | 1.09; 1.21 | 13 - 21 | |
| | | | | | | | | | | |
| SMLO | 39 | 89 | Y | 16.81 | 224.90 | 4.36 | .26 | 6.22; 6.62 | 10.94 - 19 | [58] |
| SLO | 39 | 89 | Y | 16.74 | 226.00 | 4.25 | .25 | 2.28; 12.5 | 14 - 19 | |
| | | | | | | | | | | |
| SMLO | 40 | 90 | Zr | 16.80 | 211.76 | 4.17 | .25 | 7.43; 4.48 | 12.17 - 19 | [58] |
| SLO | 40 | 90 | Zr | 16.74 | 211.00 | 4.16 | .25 | 5.90; 16.8 | 14 - 19 | |
| | | | | | | | | | | |
| SMLO | 40 | 91 | Zr | 16.65 | 183.18 | 4.35 | .26 | 1.60; 5.74 | 14 - 19 | [59] |
| SLO | 40 | 91 | Zr | 16.58 | 184.00 | 4.20 | .25 | 1.72; 2.13 | 14 - 19 | |
| | | | | | | | | | | |
| SMLO | 40 | 92 | Zr | 16.35 | 166.18 | 4.74 | .29 | 1.49; 4.38 | 14 - 19 | [59] |
| SLO | 40 | 92 | Zr | 16.26 | 166.00 | 4.68 | .29 | 1.89; 2.78 | 14 - 19 | |
| | | | | | | | | | | |
| SMLO | 40 | 94 | Zr | 16.36 | 158.83 | 5.56 | .34 | 1.40; 19.2 | 14 - 19 | [59] |
| SLO | 40 | 94 | Zr | 16.22 | 161.00 | 5.29 | .33 | 1.18; 6.80 | 14 - 19 | |
| | | | | | | | | | | |
| SMLO | 41 | 93 | Nb | 16.71 | 199.17 | 5.26 | .31 | 2.07; 17.1 | 14 - 19 | [58] |
| SLO | 41 | 93 | Nb | 16.59 | 200.00 | 5.05 | .30 | 4.20; 5.04 | 14 - 19 | |
| | | | | | | | | | | |
| SMLO | 42 | 92 | Mo | 16.88 | 163.57 | 4.00 | .24 | 17.4; 14.3 | 12.53 - 19 | [60] |
| SLO | 42 | 92 | Mo | 16.82 | 162.00 | 4.14 | .25 | 15.1; 32.6 | 14 - 19 | |
| | | | | | | | | | | |
| SMLO | 42 | 94 | Mo | 16.52 | 184.20 | 5.64 | .34 | 4.68; 7.16 | 9.68 - 19 | [60] |
| SLO | 42 | 94 | Mo | 16.36 | 185.00 | 5.50 | .34 | 6.26; 15.6 | 14 - 19 | |
| | | | | | | | | | | |
| SMLO | 42 | 96 | Mo | 16.58 | 184.90 | 6.90 | .42 | 2.86; 60.6 | 13 - 17 | [60] |
| SLO | 42 | 96 | Mo | 16.20 | 185.00 | 6.01 | .37 | 3.74; 3.90 | 13 - 17 | |
| | | | | | | | | | | |
| SMLO | 42 | 98 | Mo | 15.98 | 187.77 | 6.23 | .39 | 2.21; 5.06 | 13 - 19 | [60] |
| SLO | 42 | 98 | Mo | 15.80 | 189.00 | 5.94 | .38 | 2.10; 2.27 | 13 - 19 | |
| | | | | | | | | | | |
| SMLO | 42 | 100 | Mo | 16.04 | 169.47 | 8.23 | .51 | 1.94; 6.29 | 8.33 - 20 | [60] |
| SLO | 42 | 100 | Mo | 15.74 | 171.00 | 7.81 | .50 | 2.11; 6.24 | 12 - 20 | |
| | | | | | | | | | | |
| SMLO | 45 | 103 | Rh | 16.46 | 188.48 | 7.87 | .48 | 1.37; 12.3 | 9.16 - 19 | [61] |
| SLO | 45 | 103 | Rh | 16.16 | 191.00 | 7.26 | .45 | 1.03; 6.87 | 13 - 19 | |
| | | | | | | | | | | |
| SMLO | 47 | 107 | Ag | 16.14 | 149.43 | 7.15 | .44 | 1.77; 6.38 | 9.41 - 19 | [62] |
| SLO | 47 | 107 | Ag | 15.90 | 150.00 | 6.71 | .42 | 1.54; 5.02 | 13 - 19 | |
| | | | | | | | | | | |
| SMLO | 49 | 115 | In | 15.92 | 243.46 | 6.03 | .38 | 1.60; 112 | 13 - 18 | [61] |
| SLO | 49 | 115 | In | 15.72 | 247.00 | 5.60 | .36 | 4.20; 13.6 | 13 - 18 | |
| | | | | | | | | | | |
| SMLO | 50 | 116 | Sn | 15.70 | 268.61 | 5.32 | .34 | 1.18; 17.2 | 13 - 18 | [61] |
| SLO | 50 | 116 | Sn | 15.56 | 271.00 | 5.08 | .33 | 2.17; 4.63 | 13 - 18 | |
| | | | | | | | | | | |
| SMLO | 50 | 117 | Sn | 15.78 | 256.05 | 5.32 | .34 | 1.07; 9.91 | 13 - 18 | [61] |
| SLO | 50 | 117 | Sn | 15.64 | 259.00 | 5.04 | .32 | 1.71; 1.88 | 13 - 18 | |
| | | | | | | | | | | |
| SMLO | 50 | 118 | Sn | 15.56 | 277.24 | 5.05 | .32 | 1.02; 6.50 | 13 - 18 | [61] |
| SLO | 50 | 118 | Sn | 15.44 | 279.00 | 4.86 | .31 | 1.71; 1.30 | 13 - 18 | |
| | | | | | | | | | | |
| SMLO | 50 | 119 | Sn | 15.66 | 250.06 | 5.12 | .33 | 1.34; 17.2 | 13 - 18 | [63] |
| SLO | 50 | 119 | Sn | 15.53 | 253.00 | 4.81 | .31 | 1.34; 1.99 | 13 - 18 | |
| | | | | | | | | | | |
| SMLO | 50 | 120 | Sn | 15.51 | 283.52 | 5.29 | .34 | .93; 14.98 | 13 - 18 | [61] |
| SLO | 50 | 120 | Sn | 15.37 | 285.00 | 5.10 | .33 | 1.97; 4.30 | 13 - 18 | |
| | | | | | | | | | | |
| SMLO | 50 | 124 | Sn | 15.39 | 274.60 | 4.91 | .32 | 1.48; 2.96 | 9.41 - 18 | [61] |
| SLO | 50 | 124 | Sn | 15.28 | 276.00 | 4.80 | .31 | 2.39; 5.82 | 13 - 18 | |
| | | | | | | | | | | |
| SMLO | 52 | 124 | Te | 15.37 | 278.18 | 5.84 | .38 | 1.06; 4.72 | 12 - 19 | [64] |
| SLO | 52 | 124 | Te | 15.24 | 281.00 | 5.56 | .36 | 1.19; 1.33 | 12 - 19 | |
| | | | | | | | | | | |
| SMLO | 52 | 126 | Te | 15.28 | 293.37 | 5.66 | .37 | .68; 4.88 | 12 - 19 | [64] |
| SLO | 52 | 126 | Te | 15.17 | 296.00 | 5.44 | .36 | 1.43; 1.91 | 12 - 19 | |
| | | | | | | | | | | |
| SMLO | 52 | 128 | Te | 15.24 | 302.24 | 5.57 | .37 | 1.42; 3.75 | 12 - 19 | [64] |
| SLO | 52 | 128 | Te | 15.13 | 305.00 | 5.36 | .35 | 1.70; 1.43 | 12 - 19 | |
| | | | | | | | | | | |
| SMLO | 52 | 130 | Te | 15.22 | 317.15 | 5.23 | .34 | .68; 3.18 | 12 - 19 | [64] |
| SLO | 52 | 130 | Te | 15.12 | 320.00 | 5.03 | .33 | .96; .90 | 12 - 19 | |
| | | | | | | | | | | |
| SMLO | 55 | 133 | Cs | 15.45 | 314.44 | 5.53 | .36 | 7.73; 34.8 | 12 - 19 | [61] |
| SLO | 55 | 133 | Cs | 15.34 | 317.00 | 5.31 | .35 | 16.8; 14.4 | 12 - 19 | |
| | | | | | | | | | | |
| SMLO | 56 | 138 | Ba | 15.32 | 323.96 | 4.79 | .31 | 6.83; 13.8 | 12 - 19 | [65] |
| SLO | 56 | 138 | Ba | 15.26 | 327.00 | 4.61 | .30 | 2.47; 2.49 | 12 - 19 | |
| | | | | | | | | | | |
| SMLO | 57 | 139 | La | 15.31 | 332.29 | 4.67 | .31 | 3.68; 6.71 | 12 - 19 | [66] |
| SLO | 57 | 139 | La | 15.24 | 336.00 | 4.47 | .29 | 1.30; 1.02 | 12 - 19 | |
| | | | | | | | | | | |
| SMLO | 58 | 140 | Ce | 15.10 | 380.53 | 4.52 | .30 | 3.07; 6.82 | 12 - 19 | [64] |
| SLO | 58 | 140 | Ce | 15.04 | 383.00 | 4.41 | .29 | 1.10; 2.09 | 12 - 19 | |
| | | | | | | | | | | |
| SMLO | 58 | 142 | Ce | 14.96 | 329.96 | 5.26 | .35 | 2.18; 6.42 | 12 - 19 | [64] |
| SLO | 58 | 142 | Ce | 14.86 | 332.00 | 5.10 | .34 | 1.08; 2.01 | 12 - 19 | |
| | | | | | | | | | | |
| SMLO | 59 | 141 | Pr | 15.21 | 320.58 | 4.66 | .31 | 3.0; 16.95 | 12 - 19 | [67] |
| SLO | 59 | 141 | Pr | 15.15 | 324.00 | 4.42 | .29 | .78; 1.45 | 12 - 19 | |
| | | | | | | | | | | |
| SMLO | 60 | 142 | Nd | 15.02 | 356.99 | 4.59 | .31 | 3.34; 10.3 | 12 - 19 | [68] |
| SLO | 60 | 142 | Nd | 14.94 | 359.00 | 4.44 | .30 | .76; 2.94 | 12 - 19 | |
| | | | | | | | | | | |
| SMLO | 60 | 143 | Nd | 15.09 | 342.90 | 5.02 | .33 | 3.54; 6.91 | 12 - 19 | [68] |
| SLO | 60 | 143 | Nd | 15.01 | 349.00 | 4.75 | .32 | 1.58; 2.18 | 12 - 19 | |
| | | | | | | | | | | |
| SMLO | 60 | 144 | Nd | 15.18 | 312.83 | 5.60 | .37 | 2.49; 8.87 | 12 - 19 | [68] |
| SLO | 60 | 144 | Nd | 15.05 | 317.00 | 5.28 | .35 | 1.22; 1.48 | 12 - 19 | |
| | | | | | | | | | | |
| SMLO | 60 | 145 | Nd | 15.15 | 291.13 | 6.80 | .45 | 3.78; 5.09 | 12 - 19 | [68] |
| SLO | 60 | 145 | Nd | 14.95 | 296.00 | 6.31 | .42 | 2.73; 1.94 | 12 - 19 | |
| | | | | | | | | | | |
| SMLO | 60 | 146 | Nd | 14.89 | 305.25 | 6.10 | .41 | 2.06; 7.32 | 12 - 19 | [68] |
| SLO | 60 | 146 | Nd | 14.74 | 310.00 | 5.78 | .39 | .89; 1.40 | 12 - 19 | |
| | | | | | | | | | | |
| SMLO | 62 | 144 | Sm | 15.38 | 380.79 | 4.56 | .30 | 1.21; 2.61 | 12 - 19 | [69] |
| SLO | 62 | 144 | Sm | 15.32 | 383.00 | 4.45 | .29 | 1.25; 1.30 | 12 - 19 | |
| | | | | | | | | | | |
| SMLO | 62 | 148 | Sm | 14.92 | 337.75 | 5.19 | .35 | 2.84; 7.07 | 12 - 19 | [69] |
| SLO | 62 | 148 | Sm | 14.82 | 339.00 | 5.09 | .34 | .58; .91 | 12 - 19 | |
| | | | | | | | | | | |
| SMLO | 62 | 150 | Sm | 14.77 | 310.98 | 6.07 | .41 | .96; 2.42 | 12 - 19 | [69] |
| SLO | 62 | 150 | Sm | 14.61 | 312.00 | 5.97 | .41 | .61; .93 | 12 - 19 | |
| | | | | | | | | | | |
| SMLO | 79 | 197 | Au | 13.80 | 537.67 | 4.66 | .34 | 1.16; 1.30 | 8.08 - 17 | [42] |
| SLO | 79 | 197 | Au | 13.72 | 541.00 | 4.61 | .34 | 2.79; 9.66 | 11 - 17 | |
| | | | | | | | | | | |
| SMLO | 82 | 206 | Pb | 13.62 | 504.76 | 4.03 | .30 | 6.22; 8.86 | 10 - 17 | [70] |
| SLO | 82 | 206 | Pb | 13.59 | 514.00 | 3.85 | .28 | 2.58; 2.71 | 10 - 17 | |
| | | | | | | | | | | |
| SMLO | 82 | 207 | Pb | 13.59 | 472.39 | 4.14 | .30 | 6.72; 7.61 | 6.78 - 17 | [70] |
| SLO | 82 | 207 | Pb | 13.56 | 481.00 | 3.96 | .29 | 2.98; 6.44 | 10 - 17 | |
| | | | | | | | | | | |
| SMLO | 82 | 208 | Pb | 13.48 | 611.65 | 4.41 | .33 | 10.8; 33.3 | 7.50 - 17 | [42] |
| SLO | 82 | 208 | Pb | 13.43 | 639.00 | 4.07 | .30 | 8.94; 33.3 | 10 - 17 | |
| | | | | | | | | | | |
| SMLO | 83 | 209 | Bi | 13.49 | 507.19 | 4.29 | .32 | 8.38; 16.6 | 10 - 17 | [70] |
| SLO | 83 | 209 | Bi | 13.45 | 521.00 | 3.97 | .30 | 3.98; 3.85 | 10 - 17 | |

**Table B2.** The SLO parameters from $\chi^2$ minimization in different $\gamma$ -ray energy intervals in comparison with photoneutron shape parameters from Ref.[18]

| Model | Z | A | EL | $E_r$ *MeV* | $\sigma_r$ *mb* | $\Gamma_r$ *MeV* | *a* | $\chi^2_{in}$; $\chi^2_{out}$ | Fitting interval *MeV* | Data Ref. |
|---|---|---|---|---|---|---|---|---|---|---|
| SLO* | 20 | 40 | Ca | 20.49 | 78.27 | 6.29 | .31 | 9.90, 5.24 | 10.20 - 23 | [41] |
| SLO* | 20 | 40 | Ca | 20.16 | 94.47 | 4.50 | .22 | 2.89, 12.4 | 15 - 23 | |
| SLO | 20 | 40 | Ca | 20.00 | 95.00 | 6.00 | .30 | 18.2, 18.6 | 15 - 23 | |
| | | | | | | | | | | |
| SLO* | 29 | 65 | Cu | 16.75 | 76.04 | 6.43 | .38 | 1.14, .91 | 9.34 - 20 | [56] |
| SLO* | 29 | 65 | Cu | 16.70 | 75.19 | 6.89 | .41 | .78, 1.41 | 14 - 20 | |
| SLO | 29 | 65 | Cu | 16.70 | 75.20 | 6.89 | .41 | .78, 1.41 | 14 - 20 | |
| | | | | | | | | | | |
| SLO* | 32 | 70 | Ge | 16.83 | 91.91 | 6.76 | .40 | 2.46, 1.84 | 11.55 - 21 | [57] |
| SLO* | 32 | 70 | Ge | 16.79 | 89.38 | 7.66 | .46 | 1.30, 3.24 | 13 - 21 | |
| SLO | 32 | 70 | Ge | 16.79 | 89.40 | 7.66 | .46 | 1.30, 3.26 | 13 - 21 | |
| | | | | | | | | | | |
| SLO* | 32 | 72 | Ge | 16.71 | 109.39 | 7.00 | .42 | 1.99, 1.69 | 10.47 - 21 | [57] |
| SLO* | 32 | 72 | Ge | 16.67 | 107.20 | 7.65 | .46 | 1.25, 2.60 | 13 - 21 | |
| SLO | 32 | 72 | Ge | 16.67 | 107.00 | 7.68 | .46 | 1.25, 2.64 | 13 - 21 | |
| | | | | | | | | | | |
| SLO* | 34 | 80 | Se | 16.11 | 137.21 | 5.27 | .33 | .95, .50 | 10.20 - 17 | [57] |
| SLO* | 34 | 80 | Se | 16.13 | 136.73 | 5.44 | .34 | .47, .99 | 13 - 17 | |
| SLO | 34 | 80 | Se | 16.53 | 136.00 | 6.90 | .42 | 1.42, 2.90 | 13 - 17 | |
| | | | | | | | | | | |
| SLO* | 34 | 82 | Se | 16.66 | 148.36 | 5.86 | .35 | 1.20, 1.10 | 9.38 - 21 | [57] |
| SLO* | 34 | 82 | Se | 16.65 | 148.20 | 5.90 | .35 | 1.09, 1.21 | 13 - 21 | |
| SLO | 34 | 82 | Se | 16.65 | 148.00 | 5.91 | .35 | 1.09, 1.21 | 13 - 21 | |
| | | | | | | | | | | |
| SLO* | 39 | 89 | Y | 16.74 | 229.56 | 4.01 | .24 | 10.31, 4.19 | 10.94 - 19 | [58] |
| SLO* | 39 | 89 | Y | 16.74 | 225.97 | 4.25 | .25 | 2.27, 12.3 | 14 - 19 | |
| SLO | 39 | 89 | Y | 16.74 | 226.00 | 4.25 | .25 | 2.28, 12.5 | 14 - 19 | |
| | | | | | | | | | | |
| SLO* | 40 | 90 | Zr | 16.74 | 215.08 | 3.89 | .23 | 14.1, 8.41 | 12.17 - 19 | [58] |
| SLO* | 40 | 90 | Zr | 16.74 | 211.31 | 4.16 | .25 | 5.88, 17.0 | 14 - 19 | |
| SLO | 40 | 90 | Zr | 16.74 | 211.00 | 4.16 | .25 | 5.90, 16.8 | 14 - 19 | |
| | | | | | | | | | | |
| SLO* | 40 | 91 | Zr | 16.60 | 185.62 | 4.08 | .25 | 1.90, 1.89 | 10.80 - 19 | [59] |
| SLO* | 40 | 91 | Zr | 16.58 | 184.42 | 4.20 | .25 | 1.71, 2.13 | 14 - 19 | |
| SLO | 40 | 91 | Zr | 16.58 | 184.00 | 4.20 | .25 | 1.72, 2.13 | 14 - 19 | |
| | | | | | | | | | | |
| SLO* | 40 | 92 | Zr | 16.27 | 167.40 | 4.53 | .28 | 2.56, 2.00 | 10.02 - 19 | [59] |
| SLO* | 40 | 92 | Zr | 16.26 | 165.88 | 4.68 | .29 | 1.89, 2.75 | 14 - 19 | |
| SLO | 40 | 92 | Zr | 16.26 | 166.00 | 4.68 | .29 | 1.89, 2.78 | 14 - 19 | |
| | | | | | | | | | | |
| SLO* | 40 | 94 | Zr | 16.18 | 161.54 | 5.22 | .32 | 6.73, 1.30 | 7.85 - 19 | [59] |
| SLO* | 40 | 94 | Zr | 16.22 | 160.97 | 5.29 | .33 | 1.18, 6.80 | 14 - 19 | |
| SLO | 40 | 94 | Zr | 16.22 | 161.00 | 5.29 | .33 | 1.18, 6.80 | 14 - 19 | |
| | | | | | | | | | | |
| SLO* | 41 | 93 | Nb | 16.60 | 203.87 | 4.72 | .28 | 3.49, 5.41 | 9.04 - 19 | [58] |
| SLO* | 41 | 93 | Nb | 16.59 | 200.74 | 5.03 | .30 | 4.18, 4.94 | 14 - 19 | |
| SLO | 41 | 93 | Nb | 16.59 | 200.00 | 5.05 | .30 | 4.20, 5.04 | 14 - 19 | |
| | | | | | | | | | | |
| SLO* | 42 | 92 | Mo | 16.83 | 165.36 | 3.78 | .22 | 27.6, 19.4 | 12.53 - 19 | [60] |
| SLO* | 42 | 92 | Mo | 16.82 | 161.68 | 4.14 | .25 | 15.1, 32.3 | 14 - 19 | |
| SLO | 42 | 92 | Mo | 16.82 | 162.00 | 4.14 | .25 | 15.1, 32.6 | 14 - 19 | |
| | | | | | | | | | | |
| SLO* | 42 | 94 | Mo | 16.38 | 189.96 | 4.92 | .30 | 8.10, 11.1 | 9.68 - 19 | [60] |
| SLO* | 42 | 94 | Mo | 16.36 | 185.27 | 5.47 | .33 | 6.25, 15.2 | 14 - 19 | |
| SLO | 42 | 94 | Mo | 16.36 | 185.00 | 5.50 | .34 | 6.26, 15.6 | 14 - 19 | |
| | | | | | | | | | | |
| SLO* | 42 | 96 | Mo | 16.17 | 188.47 | 5.62 | .35 | 2.85, 3.55 | 9.14 - 17 | [60] |
| SLO* | 42 | 96 | Mo | 16.22 | 187.87 | 5.88 | .36 | 3.26, 3.23 | 13 - 17 | |
| SLO | 42 | 96 | Mo | 16.20 | 185.00 | 6.01 | .37 | 3.74, 3.90 | 13 - 17 | |
| | | | | | | | | | | |
| SLO* | 42 | 98 | Mo | 15.80 | 189.82 | 5.90 | .37 | 2.26, 2.10 | 8.60 - 19 | [60] |
| SLO* | 42 | 98 | Mo | 15.80 | 189.46 | 5.94 | .38 | 2.09, 2.26 | 13 - 19 | |
| SLO | 42 | 98 | Mo | 15.80 | 189.00 | 5.94 | .38 | 2.10, 2.27 | 13 - 19 | |
| | | | | | | | | | | |
| SLO* | 42 | 100 | Mo | 15.80 | 174.49 | 7.15 | .45 | 4.28, 3.59 | 8.33 - 20 | [60] |
| SLO* | 42 | 100 | Mo | 15.74 | 170.96 | 7.79 | .49 | 2.10, 6.13 | 12 - 20 | |
| SLO | 42 | 100 | Mo | 15.74 | 171.00 | 7.81 | .50 | 2.11, 6.24 | 12 - 20 | |
| | | | | | | | | | | |
| SLO* | 45 | 103 | Rh | 16.14 | 195.75 | 6.49 | .40 | 3.92, 2.64 | 9.16 - 19 | [61] |
| SLO* | 45 | 103 | Rh | 16.16 | 190.96 | 7.26 | .45 | 1.03, 6.93 | 13 - 19 | |
| SLO | 45 | 103 | Rh | 16.16 | 191.00 | 7.26 | .45 | 1.03, 6.87 | 13 - 19 | |
| | | | | | | | | | | |
| SLO* | 47 | 107 | Ag | 15.91 | 152.57 | 6.10 | .38 | 3.35, 2.29 | 9.41 - 19 | [62] |
| SLO* | 47 | 107 | Ag | 15.90 | 150.47 | 6.71 | .42 | 1.53, 5.06 | 13 - 19 | |
| SLO | 47 | 107 | Ag | 15.90 | 150.00 | 6.71 | .42 | 1.54, 5.02 | 13 - 19 | |
| | | | | | | | | | | |
| SLO* | 49 | 115 | In | 15.72 | 250.27 | 5.23 | .33 | 7.24, 8.01 | 9.41 - 18 | [61] |
| SLO* | 49 | 115 | In | 15.72 | 246.57 | 5.60 | .36 | 4.15, 13.0 | 13 - 18 | |
| SLO | 49 | 115 | In | 15.72 | 247.00 | 5.60 | .36 | 4.20, 13.6 | 13 - 18 | |
| | | | | | | | | | | |
| SLO* | 50 | 116 | Sn | 15.56 | 273.81 | 4.85 | .31 | 3.90, 2.77 | 9.45 - 18 | [61] |
| SLO* | 50 | 116 | Sn | 15.56 | 270.73 | 5.08 | .33 | 2.17, 4.62 | 13 - 18 | |
| SLO | 50 | 116 | Sn | 15.56 | 271.00 | 5.08 | .33 | 2.17, 4.63 | 13 - 18 | |
| | | | | | | | | | | |
| SLO* | 50 | 117 | Sn | 15.65 | 259.81 | 4.95 | .32 | 1.73, 1.82 | 8.87 - 18 | [61] |
| SLO* | 50 | 117 | Sn | 15.64 | 258.81 | 5.04 | .32 | 1.71, 1.85 | 13 - 18 | |
| SLO | 50 | 117 | Sn | 15.64 | 259.00 | 5.04 | .32 | 1.71, 1.88 | 13 - 18 | |
| | | | | | | | | | | |
| SLO* | 50 | 118 | Sn | 15.44 | 278.79 | 4.88 | .32 | 1.29, 1.71 | 10.13 - 18 | [61] |
| SLO* | 50 | 118 | Sn | 15.44 | 279.12 | 4.86 | .31 | 1.70, 1.29 | 13 - 18 | |
| SLO | 50 | 118 | Sn | 15.44 | 279.00 | 4.86 | .31 | 1.71, 1.30 | 13 - 18 | |
| | | | | | | | | | | |
| SLO* | 50 | 119 | Sn | 15.53 | 253.99 | 4.66 | .30 | 1.46, 1.51 | 9.10 - 18 | [63] |
| SLO* | 50 | 119 | Sn | 15.53 | 252.55 | 4.81 | .31 | 1.33, 1.89 | 13 - 18 | |
| SLO | 50 | 119 | Sn | 15.53 | 253.00 | 4.81 | .31 | 1.34, 1.99 | 13 - 18 | |
| | | | | | | | | | | |
| SLO* | 50 | 120 | Sn | 15.39 | 289.32 | 4.81 | .31 | 2.99, 2.94 | 9.04 - 18 | [61] |
| SLO* | 50 | 120 | Sn | 15.37 | 285.27 | 5.10 | .33 | 1.96, 4.30 | 13 - 18 | |
| SLO | 50 | 120 | Sn | 15.37 | 285.00 | 5.10 | .33 | 1.97, 4.30 | 13 - 18 | |
| | | | | | | | | | | |
| SLO* | 50 | 124 | Sn | 15.30 | 280.22 | 4.49 | .29 | 4.11, 3.72 | 9.41 - 18 | [61] |
| SLO* | 50 | 124 | Sn | 15.28 | 275.62 | 4.80 | .31 | 2.39, 5.79 | 13 - 18 | |
| SLO | 50 | 124 | Sn | 15.28 | 276.00 | 4.80 | .31 | 2.39, 5.82 | 13 - 18 | |
| | | | | | | | | | | |
| SLO* | 52 | 124 | Te | 15.25 | 282.87 | 5.43 | .36 | 1.24, 1.25 | 9.11 - 19 | [64] |
| SLO* | 52 | 124 | Te | 15.24 | 281.34 | 5.54 | .36 | 1.19, 1.32 | 12 - 19 | |
| SLO | 52 | 124 | Te | 15.24 | 281.00 | 5.56 | .36 | 1.19, 1.33 | 12 - 19 | |
| | | | | | | | | | | |
| SLO* | 52 | 126 | Te | 15.17 | 298.20 | 5.28 | .35 | 1.76, 1.53 | 9.11 - 19 | [64] |
| SLO* | 52 | 126 | Te | 15.17 | 296.14 | 5.42 | .36 | 1.43, 1.87 | 12 - 19 | |
| SLO | 52 | 126 | Te | 15.17 | 296.00 | 5.44 | .36 | 1.43, 1.91 | 12 - 19 | |
| | | | | | | | | | | |
| SLO* | 52 | 128 | Te | 15.13 | 305.28 | 5.33 | .35 | 1.43, 1.70 | 9.38 - 19 | [64] |
| SLO* | 52 | 128 | Te | 15.13 | 305.05 | 5.34 | .35 | 1.70, 1.43 | 12 - 19 | |
| SLO | 52 | 128 | Te | 15.13 | 305.00 | 5.36 | .35 | 1.70, 1.43 | 12 - 19 | |
| | | | | | | | | | | |
| SLO* | 52 | 130 | Te | 15.12 | 320.67 | 4.98 | .33 | .88, .97 | 8.57 - 19 | [64] |
| SLO* | 52 | 130 | Te | 15.12 | 319.95 | 5.02 | .33 | .96, .89 | 12 - 19 | |
| SLO | 52 | 130 | Te | 15.12 | 320.00 | 5.03 | .33 | .96, .90 | 12 - 19 | |
| | | | | | | | | | | |
| SLO* | 55 | 133 | Cs | 15.34 | 317.58 | 5.26 | .34 | 14.2, 16.8 | 9.18 - 19 | [61] |
| SLO* | 55 | 133 | Cs | 15.34 | 317.14 | 5.29 | .34 | 16.73, 14.3 | 12 - 19 | |
| SLO | 55 | 133 | Cs | 15.34 | 317.00 | 5.31 | .35 | 16.8, 14.4 | 12 - 19 | |
| | | | | | | | | | | |
| SLO* | 56 | 138 | Ba | 15.26 | 327.35 | 4.62 | .30 | 2.48, 2.47 | 8.48 - 19 | [65] |
| SLO* | 56 | 138 | Ba | 15.26 | 327.44 | 4.61 | .30 | 2.47, 2.48 | 12 - 19 | |
| SLO | 56 | 138 | Ba | 15.26 | 327.00 | 4.61 | .30 | 2.47, 2.49 | 12 - 19 | |
| | | | | | | | | | | |
| SLO* | 57 | 139 | La | 15.24 | 335.69 | 4.48 | .29 | 1.02, 1.30 | 8.90 - 19 | [66] |
| SLO* | 57 | 139 | La | 15.24 | 335.90 | 4.47 | .29 | 1.30, 1.02 | 12 - 19 | |
| SLO | 57 | 139 | La | 15.24 | 336.00 | 4.47 | .29 | 1.30, 1.02 | 12 - 19 | |
| | | | | | | | | | | |
| SLO* | 58 | 140 | Ce | 15.02 | 379.54 | 4.55 | .30 | 1.81, 1.37 | 9.11 - 19 | [64] |
| SLO* | 58 | 140 | Ce | 15.04 | 383.03 | 4.41 | .29 | 1.09, 2.06 | 12 - 19 | |
| SLO | 58 | 140 | Ce | 15.04 | 383.00 | 4.41 | .29 | 1.10, 2.09 | 12 - 19 | |
| | | | | | | | | | | |
| SLO* | 58 | 142 | Ce | 14.84 | 327.36 | 5.38 | .36 | 1.57, 1.44 | 7.76 - 19 | [64] |
| SLO* | 58 | 142 | Ce | 14.86 | 332.32 | 5.11 | .34 | 1.08, 1.97 | 12 - 19 | |
| SLO | 58 | 142 | Ce | 14.86 | 332.00 | 5.10 | .34 | 1.08, 2.01 | 12 - 19 | |
| | | | | | | | | | | |
| SLO* | 59 | 141 | Pr | 15.15 | 322.77 | 4.47 | .30 | 1.35, .83 | 9.41 - 19 | [67] |
| SLO* | 59 | 141 | Pr | 15.15 | 324.20 | 4.42 | .29 | .77, 1.43 | 12 - 19 | |
| SLO | 59 | 141 | Pr | 15.15 | 324.00 | 4.42 | .29 | .78, 1.45 | 12 - 19 | |
| | | | | | | | | | | |
| SLO* | 60 | 142 | Nd | 14.94 | 359.88 | 4.42 | .30 | 2.93, .76 | 9.45 - 19 | [68] |
| SLO* | 60 | 142 | Nd | 14.94 | 359.48 | 4.44 | .30 | .75, 2.94 | 12 - 19 | |
| SLO | 60 | 142 | Nd | 14.94 | 359.00 | 4.44 | .30 | .76, 2.94 | 12 - 19 | |
| | | | | | | | | | | |
| SLO* | 60 | 143 | Nd | 15.00 | 344.79 | 4.92 | .33 | 1.92, 1.79 | 9.31 - 19 | [68] |
| SLO* | 60 | 143 | Nd | 15.01 | 348.97 | 4.75 | .32 | 1.58, 2.15 | 12 - 19 | |
| SLO | 60 | 143 | Nd | 15.01 | 349.00 | 4.75 | .32 | 1.58, 2.18 | 12 - 19 | |
| | | | | | | | | | | |
| SLO* | 60 | 144 | Nd | 15.04 | 317.71 | 5.23 | .35 | 1.47, 1.23 | 7.95 - 19 | [68] |
| SLO* | 60 | 144 | Nd | 15.05 | 316.92 | 5.28 | .35 | 1.22, 1.48 | 12 - 19 | |
| SLO | 60 | 144 | Nd | 15.05 | 317.00 | 5.28 | .35 | 1.22, 1.48 | 12 - 19 | |
| | | | | | | | | | | |
| SLO* | 60 | 145 | Nd | 14.95 | 296.32 | 6.33 | .42 | 1.94, 2.73 | 10.13 - 19 | [68] |
| SLO* | 60 | 145 | Nd | 14.95 | 296.50 | 6.31 | .42 | 2.72, 1.94 | 12 - 19 | |
| SLO | 60 | 145 | Nd | 14.95 | 296.00 | 6.31 | .42 | 2.73, 1.94 | 12 - 19 | |
| | | | | | | | | | | |
| SLO* | 60 | 146 | Nd | 14.75 | 314.61 | 5.52 | .37 | 1.05, 1.17 | 7.95 - 19 | [68] |
| SLO* | 60 | 146 | Nd | 14.74 | 310.24 | 5.78 | .39 | .88, 1.42 | 12 - 19 | |
| SLO | 60 | 146 | Nd | 14.74 | 310.00 | 5.78 | .39 | .89, 1.40 | 12 - 19 | |
| | | | | | | | | | | |
| SLO* | 62 | 144 | Sm | 15.32 | 382.02 | 4.48 | .29 | 1.29, 1.26 | 10.50 - 19 | [69] |
| SLO* | 62 | 144 | Sm | 15.32 | 382.63 | 4.45 | .29 | 1.25, 1.30 | 12 - 19 | |
| SLO | 62 | 144 | Sm | 15.32 | 383.00 | 4.45 | .29 | 1.25, 1.30 | 12 - 19 | |
| | | | | | | | | | | |
| SLO* | 62 | 148 | Sm | 14.83 | 340.16 | 5.04 | .34 | .85, .63 | 8.33 - 19 | [69] |
| SLO* | 62 | 148 | Sm | 14.82 | 339.19 | 5.09 | .34 | .57, .90 | 12 - 19 | |
| SLO | 62 | 148 | Sm | 14.82 | 339.00 | 5.09 | .34 | .58, .91 | 12 - 19 | |
| | | | | | | | | | | |
| SLO* | 62 | 150 | Sm | 14.60 | 311.91 | 5.99 | .41 | .92, .61 | 8.06 - 19 | [69] |
| SLO* | 62 | 150 | Sm | 14.61 | 311.96 | 5.97 | .41 | .61, .92 | 12 - 19 | |
| SLO | 62 | 150 | Sm | 14.61 | 312.00 | 5.97 | .41 | .61, .93 | 12 - 19 | |
| | | | | | | | | | | |
| SLO* | 79 | 197 | Au | 13.71 | 555.25 | 4.18 | .30 | 6.13, 4.80 | 8.08 - 17 | [42] |
| SLO* | 79 | 197 | Au | 13.72 | 543.60 | 4.53 | .33 | 2.69, 8.47 | 11 - 17 | |
| SLO | 79 | 197 | Au | 13.72 | 541.00 | 4.61 | .34 | 2.79, 9.66 | 11 - 17 | |
| | | | | | | | | | | |
| SLO* | 82 | 206 | Pb | 13.58 | 509.55 | 3.92 | .29 | 2.63, 2.65 | 6.93 - 17 | [70] |
| SLO* | 82 | 206 | Pb | 13.58 | 513.51 | 3.85 | .28 | 2.57, 2.71 | 10 - 17 | |
| SLO | 82 | 206 | Pb | 13.59 | 514.00 | 3.85 | .28 | 2.58, 2.71 | 10 – 17 | |
| | | | | | | | | | | |
| SLO* | 82 | 207 | Pb | 13.58 | 491.46 | 3.76 | .28 | 5.83, 3.58 | 6.78 - 17 | [70] |
| SLO* | 82 | 207 | Pb | 13.56 | 481.13 | 3.96 | .29 | 2.98, 6.48 | 10 - 17 | |
| SLO | 82 | 207 | Pb | 13.56 | 481.00 | 3.96 | .29 | 2.98, 6.44 | 10 - 17 | |
| | | | | | | | | | | |
| SLO* | 82 | 208 | Pb | 13.36 | 656.71 | 3.66 | .27 | 24.10,17.84 | 7.50 - 17 | [42] |
| SLO* | 82 | 208 | Pb | 13.42 | 628.93 | 4.14 | .31 | 8.66, 36.2 | 10 - 17 | |
| SLO | 82 | 208 | Pb | 13.43 | 639.00 | 4.07 | .30 | 8.94, 33.3 | 10 - 17 | |
| | | | | | | | | | | |
| SLO* | 83 | 209 | Bi | 13.45 | 519.65 | 4.00 | .30 | 3.83, 3.99 | 8.01 - 17 | [70] |
| SLO* | 83 | 209 | Bi | 13.45 | 521.02 | 3.97 | .30 | 3.98, 3.84 | 10 - 17 | |
| SLO | 83 | 209 | Bi | 13.45 | 521.00 | 3.97 | .30 | 3.98, 3.85 | 10 - 17 | |